\documentclass[12pt,a4paper]{article}
\usepackage{jcappub}
\usepackage{bm}
\usepackage{indentfirst}
\usepackage{amsmath}
\usepackage{graphicx}
\usepackage{float}
\usepackage{amssymb}
\usepackage{subfigure}
\usepackage{hyperref}
\usepackage{array}
\usepackage{amsthm}
\usepackage{mathrsfs}
\usepackage{color}
\usepackage{comment}
\usepackage[normalem]{ulem}
\usepackage{url} 
\hypersetup{
	colorlinks=true,
	linkcolor=red,
	citecolor=blue,
}
\allowdisplaybreaks[2]

\begin{document}

\title{ Residual Test for the Third Gravitational-Wave Transient Catalog }

\author[a,1]{Dicong Liang,\note{Corresponding author.}}
\author[b]{Ning Dai,}
\author[a,2]{Yingjie Yang\note{Corresponding author.} }

\affiliation[a]{Department of Mathematics and Physics, School of Biomedical Engineering, Southern Medical University, Guangzhou, 510515, China}
\affiliation[b]{Department of Mathematics and Physics, Suzhou Polytechnic University, Suzhou, Jiangsu 215000, China}

\emailAdd{dcliang@smu.edu.cn}
\emailAdd{daining@jssvc.edu.cn}
\emailAdd{yyj@smu.edu.cn}


\abstract{ The residual test is commonly used to check the agreement between the gravitational wave signal and the theoretical waveform template. 
The basic idea of the residual test is to subtract the best-fit waveform from the data and then check whether the remaining data (i.e., the residuals)  are consistent with the instrumental noise or not.  
We apply the Kolmogorov-Smirnov test, the Anderson-Darling test and  the chi-squared test as goodness-of-fit test to examine the residuals of events in the third gravitational-wave transient catalog and find no statistically significant deviation from the noise. 
Although our method is sensitive only to the loud events, it does not rely on the cross-correlation between detectors. 
A single-detector event suffices for our residual analysis, and the test is simple and computationally inexpensive.
}

\maketitle

\section{Introduction}

Recently, the LIGO-Virgo-KAGRA collaboration released the fourth gravitational-wave transient catalog (GWTC-4) \cite{LIGOScientific:2025hdt}. 
Before that, there were already 90 high-confident events reported in the previous catalogs \cite{LIGOScientific:2018mvr,LIGOScientific:2020ibl,LIGOScientific:2021usb,KAGRA:2021vkt}.
The increasing number of gravitational wave (GW) detections enable us to investigate the fields of astrophysics and fundamental physics deeply. 

If there is a single GW signal immersed in the detector noise, the data can be expressed as the sum of the signal and the noise,
\begin{align}
    d(t) = n(t) +s(t) . 
\end{align}
One can use the matched-filtering technique to estimate the parameters of the GW signal with a specific waveform model.
After subtracting the best-fit template $h(t)$ from the data, the residual is given by 
\begin{align}
    r(t) = d(t) -h(t) = n(t) + s(t)-h(t).
\end{align}
If the template accurately represents the signal, only noise remains in the residual. 
This means that the residual will be statistically consistent with the instrumental noise.
In most cases, the noise in LIGO detectors can be approximated as Gaussian, stationary noise over limited time scales
and frequency ranges \cite{LIGOScientific:2019hgc,LIGO:2021ppb,LIGO:2024kkz}.
From this perspective, deviations of the residuals from the Gaussian noise indicate either the presence of glitches or poor fit of the template.
If the inconsistency between the residuals and Gaussian noise is caused by glitches, they can be mitigated using several methods (e.g., \cite{Powell:2018csz,Pankow:2018qpo,Cornish:2020dwh,Merritt:2021xwh,Kwok:2021zny,Hourihane:2022doe,Davis:2022ird,Mohanty:2023mjn,Narola:2024qdh}),
while the systematic error in waveform modeling can inform us more about astrophysics and theoretical physics.

To date, there are plenty of waveform templates, but most can only describe compact binary coalescence in a quasi-circular orbit in vacuum.
Thus, waveform models that incorporate orbital eccentricity (e.g., \cite{Hinderer:2017jcs,Cao:2017ndf,Liu:2021pkr,Nagar:2021gss,Khalil:2021txt,Islam:2021mha,Ramos-Buades:2021adz,Wang:2023ueg,LIGOScientific:2023lpe,Gupte:2024jfe,Trenado:2025ccf}), environmental effects from the gaseous or dark matter backgrounds (e.g., \cite{Barausse:2007dy,Dai:2021olt,Cole:2022ucw,Garg:2022nko,Dai:2023cft,Zwick:2024yzh,Zwick:2025wkt}), and three-body interaction (e.g., \cite{Torigoe:2009bw,Dmitrasinovic:2014lha,Bonetti:2017hnb,Gupta:2019unn,Chandramouli:2021kts}) are still in development.
If two or more GW signals overlap with each other but only the template of the loudest one is subtracted, the remaining signals introduce deviations from Gaussian noise in the residuals \cite{Samajdar:2021egv,Relton:2021cax,Antonelli:2021vwg,Wang:2023ldq,Wang:2025aqk}.
Last but not least, most mature waveform templates are developed within the framework of General Relativity (GR).
Thus, a poor fit can originate from the effects beyond GR, such as dispersion or birefringence during propagation (e.g., \cite{Kostelecky:2016kfm,Mewes:2019dhj,Shao:2020shv,Wang:2021ctl,Zhao:2022pun,Haegel:2022ymk,Zhu:2023rrx}), extra polarizations (e.g., \cite{Liang:2017ahj,Hou:2017bqj,Soudi:2018dhv,Gong:2018cgj,Capozziello:2019msc,Liang:2022hxd,Dong:2023bgt,Dong:2024zal}), and so on.
From this perspective, the residual test provides a broad check for any of the aforementioned features which are not modeled in the waveform template.

The first residual test was applied to the first detected event GW150914, using the {\tt BAYESWAVE} algorithm \cite{Cornish:2014kda} in Ref.~\cite{LIGOScientific:2016lio}.
It was subsequently applied to test residuals in the later GW events \cite{LIGOScientific:2019fpa,LIGOScientific:2020zkf,LIGOScientific:2020ufj,LIGOScientific:2020tif,LIGOScientific:2021sio}. 
Residual tests based on Pearson cross-correlation were performed in Refs.~\cite{Green:2017voq,Nielsen:2018bhc,Marcoccia:2020rag}.
Since these works were based on the correlation between detectors, at least two detectors are required.

Previously, \citet{Liang:2020rt} proposed using the GW amplitude in the time domain and q-transformed energies as statistics, and applied chi-squared goodness-of-fit test to the residuals of events in GWTC-1. They found no significant deviation from GR or other effect that
not included in the waveform model. 
Since the time-domain amplitude is insensitive to residual deviations, we use only the q-transformed energies as test statistics in this work.
The q-transform, a modification of the standard Fourier transform, is commonly used to visualize GW signals and to search for excess signal power in the time-frequency plane in the noisy data \cite{Chatterji:2004qg,chatterji2005search,Blackburn:2008ah,Vazsonyi:2022jul}.
Since the normalized q-transformed energies of white noise are expected to follow the exponential distribution \cite{chatterji2005search,Vazsonyi:2022jul}, we use them as statistics to perform residual tests.
Specifically, we extend the previous study \cite{Liang:2020rt} by introducing two additional goodness-of-fit tests.
In addition to the chi-squared test ($\chi^2$ test), we also introduce the Kolmogorov-Smirnov (KS) test and the Anderson-Darling (AD) test  to improve the robustness of our results.
In this work, we focus on the events in GWTC-3 \cite{KAGRA:2021vkt}, and we plan to apply the analysis to GWTC-4 in the upcoming future work.

The paper is organized as follows.
We discuss the waveform subtraction procedure and the statistical properties of the residuals in Sec.~\ref{sec2}.
Sec.~\ref{sec3} presents three residual tests in detail, which are the KS test, AD test, and $\chi^2$ test.
Finally, we summarize and discuss the results in Sec.~\ref{sec4}.

\section{Waveform Subtraction}
\label{sec2}

Advanced LIGO \cite{LIGOScientific:2014pky} consists of two interferometric detectors in the United States: one in Hanford, Washington, and the other in Livingston, Louisiana -- referred to as “H1” and “L1” respectively.
The strain data from these two detectors and the posterior samples from parameter estimation are obtained from the Gravitational Wave Open Science Center (GWOSC) \cite{Trovato:2019liz} \footnote{\url{https://gwosc.org/eventapi/html/GWTC-3-confident/}}.
Although Advanced Virgo \cite{VIRGO:2014yos} contributed to the GWTC-3 detections, but we do not use the Virgo data in this work since most events are weak (low-SNR) in Virgo \cite{KAGRA:2021vkt}. 
To generate the waveform template, we use the phenomenological model  {\tt IMRPhenomXPHM} \cite{Pratten:2020ceb}, which incorporates multipoles beyond the dominant quadrupole in the precessing frame.
Specifically, we choose a 32-second-long data segment with the sampling frequency 4k Hz for each event.
We then select the parameter set that maximizes the likelihood to generate the template.

\begin{figure*}[t]
\centering
\includegraphics[width=0.95\textwidth]{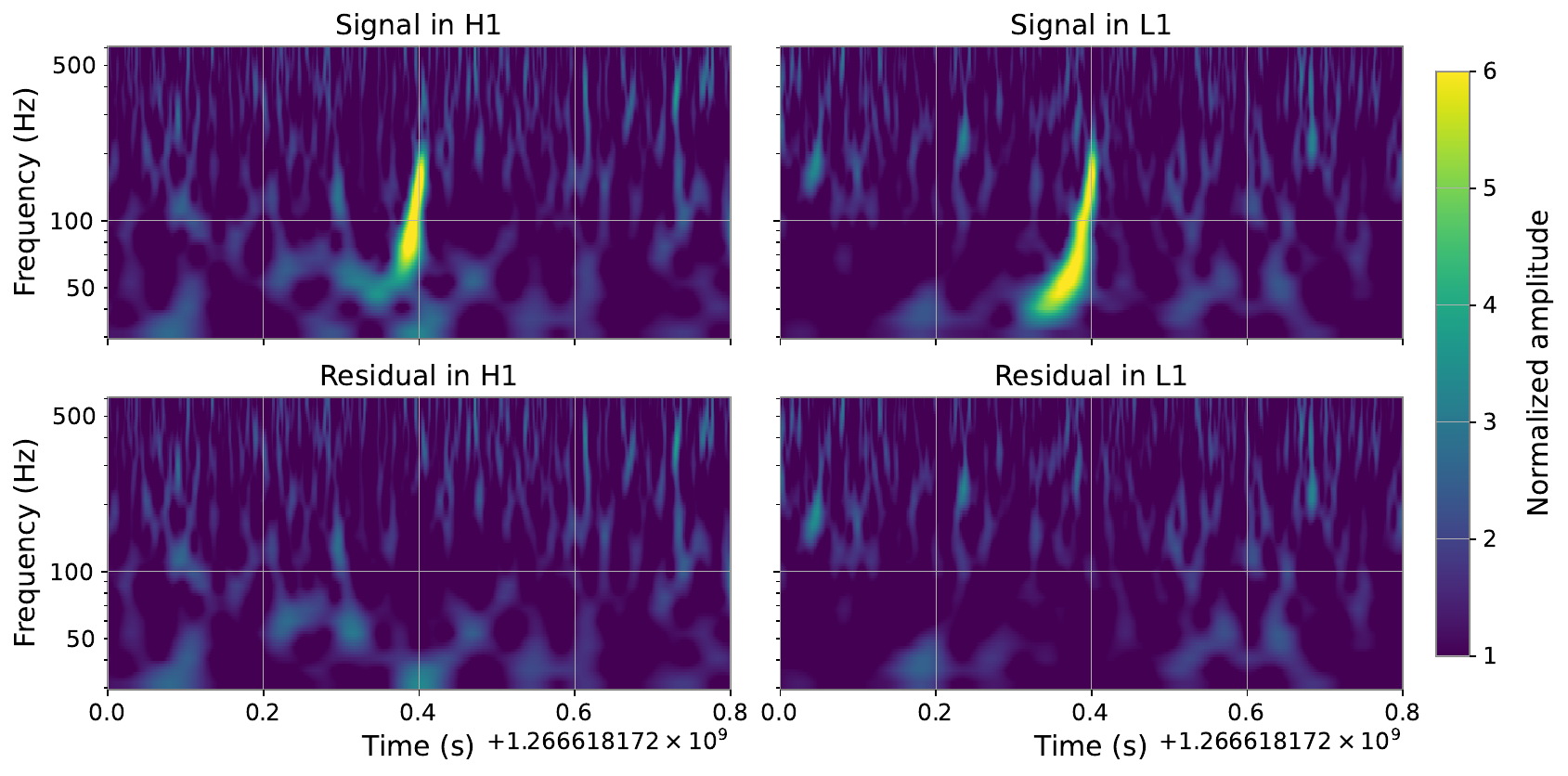}
\caption{The visualization of the signals and the residuals of event GW200224\_222234 in the time-frequency plane.}
\label{fig:event}
\end{figure*}

Here, we take the event GW200224\_222234 as an example.
In Fig.~\ref{fig:event}, we show the energy distribution in the time-frequency plane before and after subtracting the best-fit template from the strain data of the two detectors.
As shown, two bright tracks appear in the signal plots, representing the chirp signals.
However, after subtraction, the signals disappear completely, and the residuals appear consistent with noise.

To quantify the consistency, we use the q-transform energies as the statistic.
In general, the continuous q-transform of time series $x(t)$ is given by \cite{Chatterji:2004qg, chatterji2005search}
\begin{equation}
    X(\tau,f)=\int^{+\infty}_{-\infty}x(t) w(t-\tau,f,Q)e^{-i2\pi f t}dt,
\end{equation}
where $w(t-\tau,f,Q)$ is a time-domain window centered at time $\tau$, with a duration proportional to Q and inversely proportional to the frequency $f$.
For the discrete GW data, the corresponding q-transform becomes
\begin{equation}
    X[m,k]=\sum_{n=0}^{N-1} x[n] w[n-m,k] e^{-2\pi i nk/N},
\end{equation}
where $m$ and $k$ are discrete indices corresponding to time $\tau$ and frequency $f$, respectively.
It has been shown that, for stationary white noise, the normalized q-transformed energy, i.e., $y=|X|^2/ \langle |X|^2 \rangle$ follows an exponential distribution \cite{chatterji2005search,Vazsonyi:2022jul}. 
In other word, the probability density function is given by
\begin{equation}
    f(y)=e^{-y}.
    \label{exp_pdf}
\end{equation}

\begin{figure*}[t]
\centering
\includegraphics[width=0.9\textwidth]{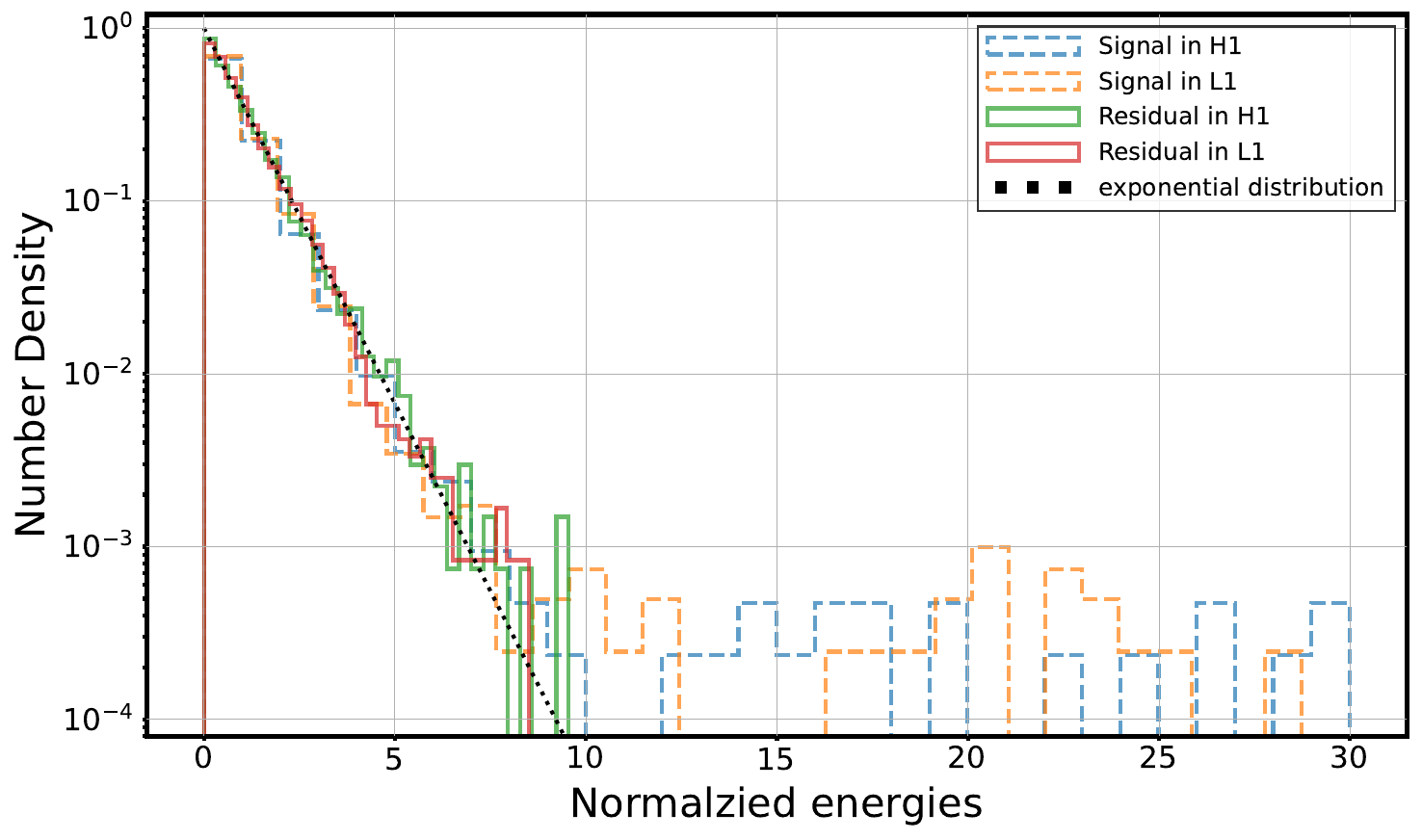}
\caption{Distribution of the normalized q-transformed energies of the signals and residuals.
The exponential curve is shown as a black dotted line.}
\label{fig:hist}
\end{figure*}

We first whiten the data and then adopt the {\tt q-gram}
module in {\tt GWpy} \cite{Macleod:2021goi} to compute the q-transformed energies.
Specifically, we select only the energies computed from one-segment data, which begins at 0.8 second before the merger time and ends at 0.2 second after.
The frequency range spans from 30 Hz to 500 Hz, which contains most of the signal power.
The distribution of normalized q-transformed energies is shown in Fig.~\ref{fig:hist}. 
Visually, the histograms of the residuals from both detectors are consistent with the expected exponential distribution.
However, due to the presence of loud signals, the histograms of the signal-containing data exhibit long tails.
In the following section, we perform three goodness-of-fit tests using these normalized q-transformed energies.

\section{Goodness of Fit Tests}
\label{sec3}
\subsection{Kolmogorov-Smirnov Test}
\label{sec3.1}

\begin{figure*}[t]
\centering
\includegraphics[width=0.95\textwidth]{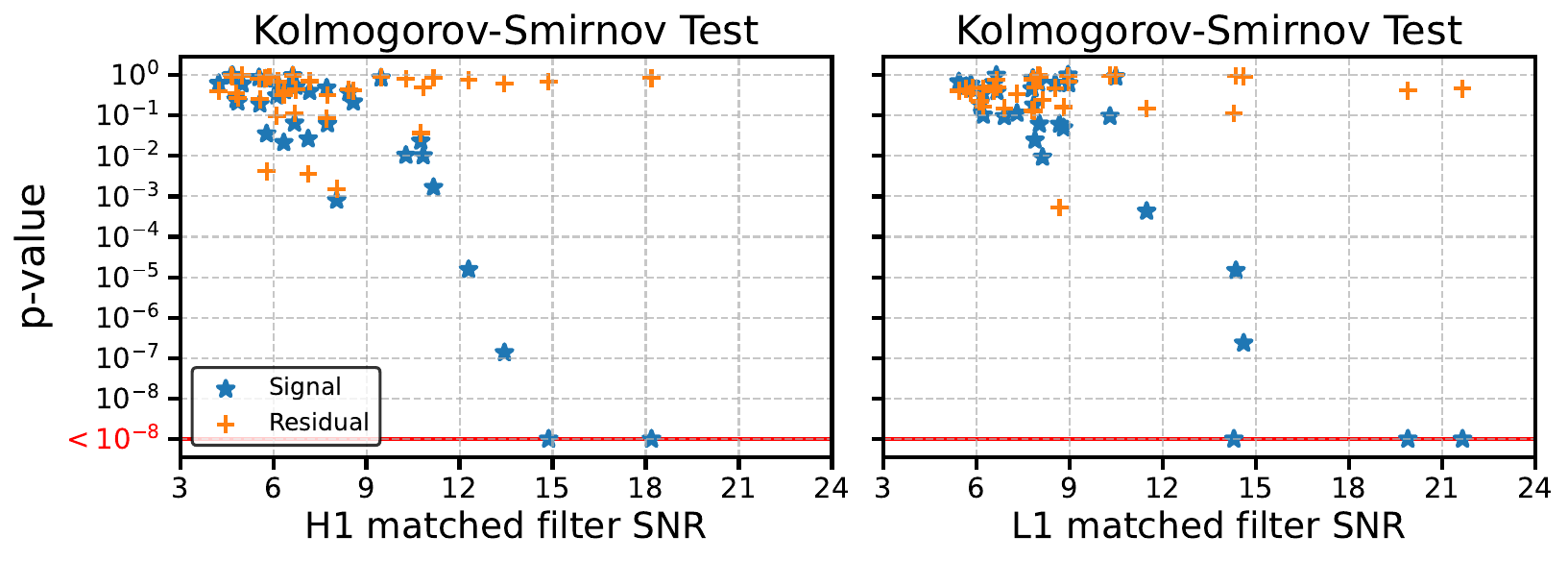}
\caption{p-values from the Kolmogorov-Smirnov test applied to the signals and residuals. 
The results of the signals and the residuals are indicated by blue asterisks (*) and orange plus signs (+), respectively. 
For clarity, p-values below $10^{-8}$ are displayed at a common height. }
\label{fig:kstest}
\end{figure*}

The KS test is a commonly used nonparametric test of the equality of one-dimensional probability distributions, which can also serve as a goodness-of-fit test.
The KS statistic is given by the largest absolute difference between the empirical cumulative distribution of the samples $F_N(y)$, and the expected cumulative distribution $F(y)$ across all $y$ values \cite{an1933sulla}, that is,
\begin{align}
    D_N(y) = \sup_y|F_N(y)-F(y)|.
\end{align}
Here, the corresponding reference cumulative distribution function for $y$ is given by
\begin{equation}
    F(y)=1-\text{exp}(-y).
    \label{exp_cdf}
\end{equation}
Under the null hypothesis, the distribution of $\sqrt{N}D_N$ converges to the Kolmogorov distribution as $N\to \infty$ \cite{an1933sulla,smirnov1948table}.
Thus we can estimate the p-value as
\begin{align}
    p = 2 \sum_{j=1}^{\infty} (-1)^{j-1} e^{-2j^2 (\sqrt{N}D_N)^2 }.
\end{align}
In practice, we adopt the {\tt kstest} module in {\tt scipy} \cite{virtanen2020scipy} to compute the KS statistic and the corresponding p-value for the residuals.
For comparison, we also apply the KS test to the whitened strain data containing signals.
We show the p-values of the KS tests in Fig.~\ref{fig:kstest}. 
Results for the data containing signals and the residuals are marked with blue ``*" and orange ``+", respectively.

For the residuals or the weak signals, the p-values are relatively large ($\gtrsim 10^{-4}$), indicating that they are statistically indistinguishable from noise.
In contrast, for the loud signals (SNR $\gtrsim 12$), their q-transformed energies deviate significantly from the exponential distribution due to their strong power.
Thus, we obtain extremely small p-values when applying the KS test to these signals.
To better visualize the data in the same figure, all data points with p-values smaller than $10^{-8}$ are plotted at the same height.
We note the presence of glitches near the event GW191113\_071753 in the low-frequency band.
Thus, for this special event, we only consider the q-transformed energies in frequency range 50Hz to 500Hz.

\subsection{Anderson-Darling Test}
\label{sec3.2}

AD test is also commonly used as a goodness-of-fit test for distributions.
The statistic of AD test is given by \cite{anderson1954test}
\begin{align}
    A^2 = -N - \sum_{i=1}^N  \frac{2i-1}{N} 
    \left[ \ln(F(Y_i))-\ln(1- F(Y_{n+1-i}))  \right], 
\end{align}
where $Y_i$ denotes the data sorted in increasing order.
We first compute $A^2$ for both the signals and the residuals.
Then, we use $10^8$ simulations to approximate the distribution of $A^2$ under the null hypothesis.
Finally, the p-value is computed as,
\begin{align}
    p=P(A^2_\text{sim} \ge A^2_\text{ data}| \text{simulations}),
\end{align}
where the probability is estimated from the simulated distribution.

\begin{figure*}[t]
\centering
\includegraphics[width=0.95\textwidth]{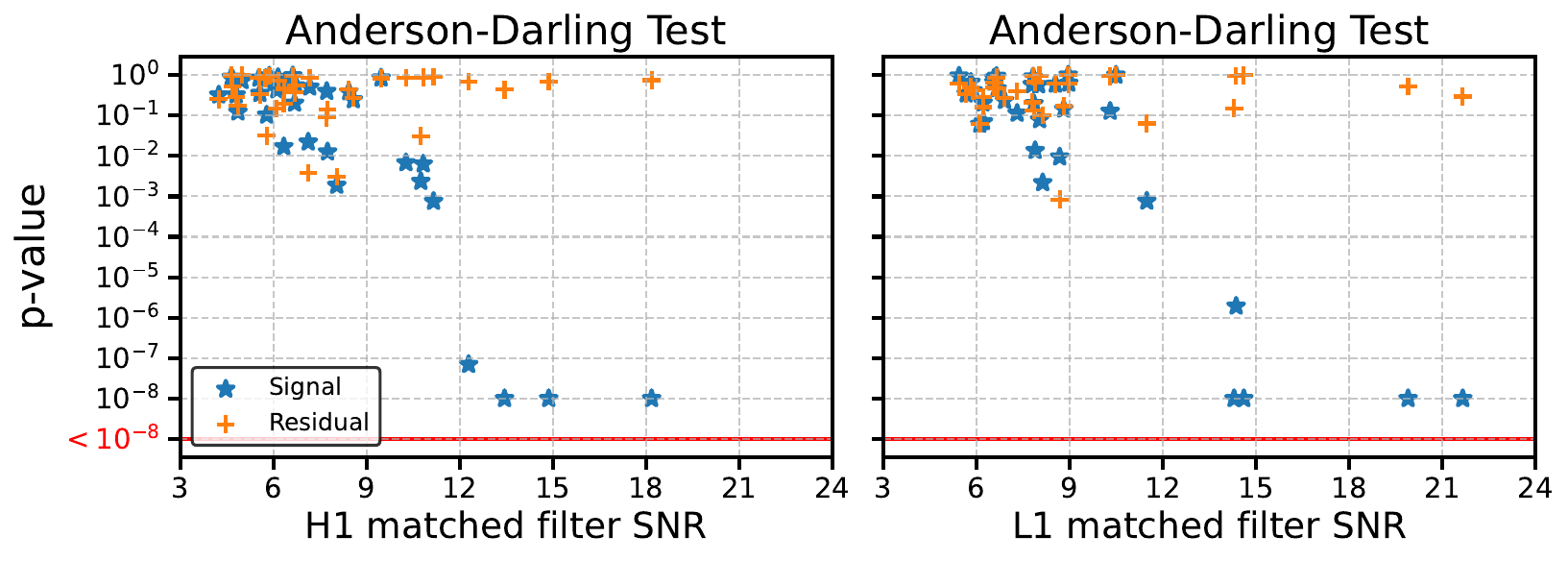}
\caption{The p-values from the Anderson-Darling test applied to the signals and the residuals. 
Symbol conventions and visualization settings follow those in Fig.~\ref{fig:kstest}. }
\label{fig:adtest}
\end{figure*}

The results are shown in Fig.~\ref{fig:adtest}. 
The p-value distributions for the signals and the residuals are quite similar to those obtained from the KS test.
The results of signals and residuals become well separated when the signals are loud.

\subsection{Chi-squared Test}
\label{sec3.3}

Lastly, we apply the Pearson's chi-squared test to the signals and the residuals.
We first group the data into $n_b$ bins, and then compute the statistic $\chi^2$ as follows \cite{pearson1900x}:
\begin{equation}
    \chi^2\equiv\sum^{n_b}_{i=1}\frac{[O_i-E_i]^2}{E_i},
\label{chi2}
\end{equation}
where $O_i$ is the observed count in the $i$th bin, and $E_i$ is the expected count according to the distribution in the null hypothesis.
We choose $n_b=9$ throughout this work.
After obtaining the $\chi^2$ values of the signals and the residuals,
we compute the corresponding p-values directly using the $\chi^2$ distribution.

\begin{figure*}[t]
\centering
\includegraphics[width=0.95\textwidth]{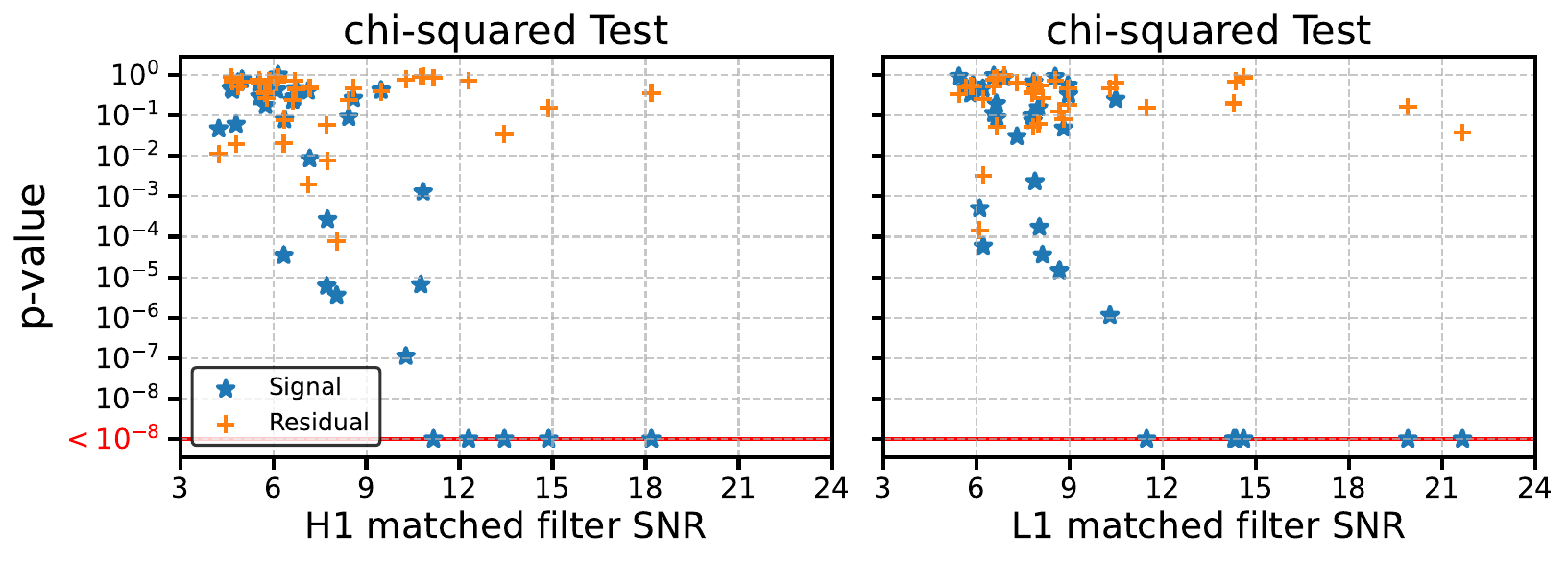}
\caption{The p-values from the chi-squared test applied to the signals and the residuals.  
Symbol conventions and visualization settings follow those in Fig.~\ref{fig:kstest}.}
\label{fig:chi2test}
\end{figure*}

The results are consistent with those of the previous two tests and are displayed in Fig.~\ref{fig:chi2test}.
Overall, no significant deviations of the residuals from the noises are found.

\section{Summary and Discussion}
\label{sec4}

In this work, we subtract the maximum-likelihood waveform from the strain data for events in GWTC-3,
and then apply residual test to assess whether the residuals are statistically consistent with instrumental noise.
We use the normalized q-transformed energies of the residuals as the test statistic and apply three goodness-of-fit tests,
including the KS test, the AD test, and the $\chi^2$ test.
We estimate the p-values under the null hypothesis that the test statistic follows an exponential distribution.
The results show that the p-values for most residuals are relatively large ($>10^{-3}$), indicating consistency with instrumental noise (the exceptions will be discussed later).
In other words, the waveform templates are sufficiently accurate for the sensitivity of current detectors.

\begin{figure*}[t]
\centering
\includegraphics[width=0.95\textwidth]{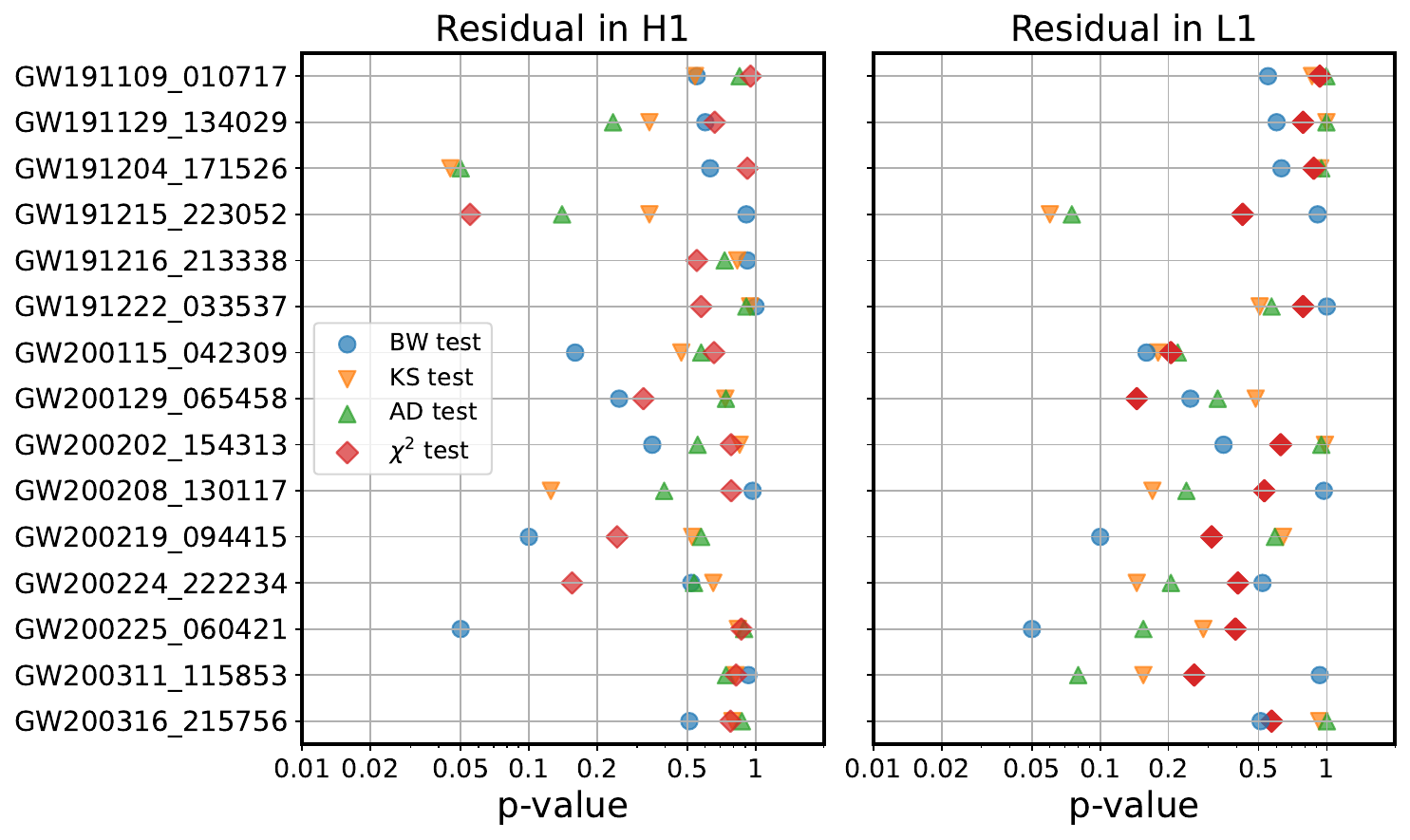}
\caption{ The p-values for residuals computed against the empirical noise distribution in the Kolmogorov-Smirnov (KS), Anderson-Darling (AD), and chi-squared ($\chi^2$) tests.
The results of  {\tt BAYESWAVE} (BW) test in Ref.~\cite{LIGOScientific:2021sio} are shown for comparison.  }
\label{fig:comparison}
\end{figure*}

For direct comparison with the results in Ref.~\cite{LIGOScientific:2021sio}, we apply the three tests to 200 randomly selected time segments within a time window of 4096 seconds symmetrically centered on the merger time for each event.
Then, we can evaluate the p-values relative to the empirical distribution of noise background, i.e., $p=P( D_N^{\rm noise} \ge D_N^{\rm residual}  |{\rm noise})$, $p=P( A^2_{\rm noise} \ge A^2_{\rm residual}  |{\rm noise})$ and $p=P( \chi^2_{\rm noise} \ge \chi^2_{\rm residual}  |{\rm noise})$.
The results of the three tests are shown in Fig.~\ref{fig:comparison}. 
The {\tt BAYESWAVE} test was applied in Ref.~\cite{LIGOScientific:2021sio} and corresponding results are denoted with blue circle in Fig.~\ref{fig:comparison}. 
Note that the LIGO Livingston detector was offline during GW191216\_213338; consequently, only data from LIGO Hanford and Virgo were used for this event in Ref.~\cite{LIGOScientific:2021sio}.
All p-values from our three tests exceed 0.04, indicating no significant deviation of residuals from instrumental noise.
Overall, our results agree with Ref.~\cite{LIGOScientific:2021sio}.
We note that our methods are less sensitive than the {\tt BAYESWAVE} test, since we only consider single-detector data separately and ignore the cross-correlation between different detectors.

Four p-values for residuals fall below $10^{-3}$ in Figs.~\ref{fig:kstest}, ~\ref{fig:adtest}, and ~\ref{fig:chi2test}. 
These are listed in Table~\ref{tab:outlier} and marked with wavy underline. 
For event GW191215\_223052 in Livingstion, the p-values for the residual in KS test and AD test are $<10^{-3}$, while the p-value of $\chi^2$ test, 0.13, is relatively large. 
For event GW191230\_180458 in Hanford and event GW200220\_124850 in Livingston, the p-values of residuals in $\chi^2$ test are $<10^{-3}$, while the p-values of KS test and AD test are relatively large.
Apparent discrepancies between individual analytic p-values arise because these different tests are sensitive to different features of the distribution. 
The $\chi^2$ test emphasizes local bin-wise deviations, and KS and AD tests focus on the global agreement of the empirical cumulative distribution with the null.
Thus, it is hard to claim significant deviation from only one of these tests.
We also compute the p-values of these residuals with respect to the 200 noise samples near the events, to account for the possible imperfection of the noise. These p-values relative to the noise background, are denoted by $p_\text{bkg}$ in the Table~\ref{tab:outlier}.
All observed test statistics fall within the typical range expected from those noise samples, indicating that none of the residuals show significant deviation.

\begin{table}[!htbp]
  \vspace{20pt}
  \centering
  \begin{tabular}{p{2cm}p{4cm}p{4cm}p{4cm}}
  \hline
  events & GW191215\_223052 & GW191230\_180458 & GW200220\_124850  \\
  \hline
  SNR &  8.7 (L1) &  8.0 (H1)   &   6.1 (L1)  \\
  \hline
  $p_{\rm KS}$ & \uwave{$5.4\times 10^{-4}$} & $1.5\times 10^{-3}$  &   0.22    \\
  $p_{\rm AD}$ & \uwave{$8.3\times 10^{-4}$} & 0.39  &  0.063  \\
  $p_{\rm \chi^2}$ & 0.13 & \uwave{$7.7\times 10^{-5}$}   &   \uwave{$1.4\times 10^{-4}$}    \\
  \hline
  $p_{\rm  bkg}$ &  0.060 (KS)  &  0.020 ($\chi^2$)  &   0.045  ($\chi^2$)  \\
                   &  0.075 (AD)  &                   &      \\
  \hline
  \end{tabular}
    \caption{Residuals for which at least one p-value is less than $10^{-3}$. The p-value with respect to the noise background is denoted by $p_\text{bkg}$. }  
  \label{tab:outlier}
\end{table}

We also apply the same goodness of fit tests to the data containing signals. 
For weak signals, the test results are indistinguishable from those of the residuals and instrumental noise.
In contrast, for strong signals (SNR $\gtrsim$ 12), the p-values are extremely small.
From this perspective, our residual tests are sensitive only to loud events.
Although loud events are not abundant in the era of second-generation terrestrial detectors, while it is expected to change with next-generation detectors.
For the next generation detectors such as the Einstein Telescope \cite{Punturo:2010zz} and the Cosmic Explorer \cite{Reitze:2019iox}, 
the sensitivity is projected to improve by an order of magnitude \cite{Hild:2010id}, 
and the SNR of GW events may reach several hundred.
Thus, our residual tests will become a powerful tool in future terrestrial GW detection.

Note that we consider events GW191216\_213338, GW200112\_155838 and  GW200302\_\allowbreak015811 as single-detector events, 
since only one LIGO detector was in operation when the gravitational waves arrived,
and the SNR was $<4.0$ for Virgo in each case \cite{KAGRA:2021vkt}.
The results of the tests are shown in the Table~\ref{tab:single}. 
The GW191216\_213338 and GW200112\_\allowbreak155838 events  signals are so strong that the p-values are extremely small for the data containing signals,
while residuals are consistent with noise in these tests.
As for the event GW200302\_015811, we can still distinguish the signal and the residual explicitly with $\chi^2$ test but KS and AD test fail.
For events in which we must rely solely on the statistical properties of data from a single detector, 
previous methods based on cross-correlation are not applicable.
Thus, our residual tests provide a complementary method for GW signal validation.

\begin{table}[!htbp]
  \vspace{20pt}
  \centering
  \begin{tabular}{p{2cm}p{4cm}p{4cm}p{4cm}}
  \hline
  events & GW191216\_213338 & GW200112\_155838 & GW200302\_015811  \\
  \hline
  SNR & 18.2 (H1) & 19.9 (L1)   &   11.2 (H1)  \\
  \hline
  $p_{\rm KS-s}$ &  $< 10^{-8}$  &  $< 10^{-8}$    &   $1.66\times 10^{-3}$    \\
  $p_{\rm KS-r}$ & 0.84 & 0.41   &   0.85     \\
  \hline
  $p_{\rm AD-s}$ &  $< 10^{-8}$ & $< 10^{-8}$  &   $7.54 \times 10^{-4}$    \\
  $p_{\rm AD-r}$ & 0.74 & 0.52   &   0.91     \\
  \hline
  $p_{\rm \chi^2-s}$ &  $< 10^{-8}$  &  $< 10^{-8}$   &    $< 10^{-8}$   \\
  $p_{\rm \chi^2-r}$ & 0.36 & 0.16   &   0.84    \\
  \hline
  \end{tabular}
    \caption{The p-value of the tests for the single-detector events. The first row shows the matched-filter SNR of the event in the corresponding detector. The second to the seventh rows show the p-value for the KS test, AD test and $\chi^2$ test respectively. The subscript `s' denotes signal and the subscript `r' denotes residual.  }
  \label{tab:single}
\end{table}

Last but not least, our method is straightforward, intuitive, and computationally inexpensive.

\acknowledgments 
We thank the anonymous referee for constructive comments that improved this work.
This work was supported by the National Natural Science Foundation of China (Grant Nos. 12405065, 12305066, and 12465013). 
Ning Dai gratefully acknowledges support from the Research Center for Intelligent Data Inference and Secure Decision-Making (Grant No. KY2025PT10) and the Fundamental Research Fund (Grant No. 202405000023) of Suzhou Polytechnic University.


\begin{thebibliography}{88}%
	\makeatletter
	\providecommand \@ifxundefined [1]{%
		\@ifx{#1\undefined}
	}%
	\providecommand \@ifnum [1]{%
		\ifnum #1\expandafter \@firstoftwo
		\else \expandafter \@secondoftwo
		\fi
	}%
	\providecommand \@ifx [1]{%
		\ifx #1\expandafter \@firstoftwo
		\else \expandafter \@secondoftwo
		\fi
	}%
	\providecommand \natexlab [1]{#1}%
	\providecommand \enquote  [1]{``#1''}%
	\providecommand \bibnamefont  [1]{#1}%
	\providecommand \bibfnamefont [1]{#1}%
	\providecommand \citenamefont [1]{#1}%
	\providecommand \href@noop [0]{\@secondoftwo}%
	\providecommand \href [0]{\begingroup \@sanitize@url \@href}%
	\providecommand \@href[1]{\@@startlink{#1}\@@href}%
	\providecommand \@@href[1]{\endgroup#1\@@endlink}%
	\providecommand \@sanitize@url [0]{\catcode `\\12\catcode `\$12\catcode
		`\&12\catcode `\#12\catcode `\^12\catcode `\_12\catcode `\%12\relax}%
	\providecommand \@@startlink[1]{}%
	\providecommand \@@endlink[0]{}%
	\providecommand \url  [0]{\begingroup\@sanitize@url \@url }%
	\providecommand \@url [1]{\endgroup\@href {#1}{\urlprefix }}%
	\providecommand \urlprefix  [0]{URL }%
	\providecommand \Eprint [0]{\href }%
	\providecommand \doibase [0]{http://dx.doi.org/}%
	\providecommand \selectlanguage [0]{\@gobble}%
	\providecommand \bibinfo  [0]{\@secondoftwo}%
	\providecommand \bibfield  [0]{\@secondoftwo}%
	\providecommand \translation [1]{[#1]}%
	\providecommand \BibitemOpen [0]{}%
	\providecommand \bibitemStop [0]{}%
	\providecommand \bibitemNoStop [0]{.\EOS\space}%
	\providecommand \EOS [0]{\spacefactor3000\relax}%
	\providecommand \BibitemShut  [1]{\csname bibitem#1\endcsname}%
	\let\auto@bib@innerbib\@empty
	\bibitem [{\citenamefont {Abac}\ \emph {et~al.}(2025)\citenamefont {Abac} \emph
		{et~al.}}]{LIGOScientific:2025hdt}%
	\BibitemOpen
	\bibfield  {author} {\bibinfo {author} {\bibfnamefont {A.~G.}\ \bibnamefont
			{Abac}} \emph {et~al.} (\bibinfo {collaboration} {LIGO Scientific, KAGRA,
			VIRGO}),\ }\href {\doibase 10.3847/2041-8213/ae0c06} {\bibfield  {journal}
		{\bibinfo  {journal} {Astrophys. J. Lett.}\ }\textbf {\bibinfo {volume}
			{995}},\ \bibinfo {pages} {L18} (\bibinfo {year} {2025})},\ \Eprint
	{http://arxiv.org/abs/2508.18080} {arXiv:2508.18080 [gr-qc]} \BibitemShut
	{NoStop}%
	\bibitem [{\citenamefont {Abbott}\ \emph
		{et~al.}(2019{\natexlab{a}})\citenamefont {Abbott} \emph
		{et~al.}}]{LIGOScientific:2018mvr}%
	\BibitemOpen
	\bibfield  {author} {\bibinfo {author} {\bibfnamefont {B.~P.}\ \bibnamefont
			{Abbott}} \emph {et~al.} (\bibinfo {collaboration} {LIGO Scientific,
			Virgo}),\ }\href {\doibase 10.1103/PhysRevX.9.031040} {\bibfield  {journal}
		{\bibinfo  {journal} {Phys. Rev. X}\ }\textbf {\bibinfo {volume} {9}},\
		\bibinfo {pages} {031040} (\bibinfo {year} {2019}{\natexlab{a}})},\ \Eprint
	{http://arxiv.org/abs/1811.12907} {arXiv:1811.12907 [astro-ph.HE]}
	\BibitemShut {NoStop}%
	\bibitem [{\citenamefont {Abbott}\ \emph
		{et~al.}(2021{\natexlab{a}})\citenamefont {Abbott} \emph
		{et~al.}}]{LIGOScientific:2020ibl}%
	\BibitemOpen
	\bibfield  {author} {\bibinfo {author} {\bibfnamefont {R.}~\bibnamefont
			{Abbott}} \emph {et~al.} (\bibinfo {collaboration} {LIGO Scientific,
			Virgo}),\ }\href {\doibase 10.1103/PhysRevX.11.021053} {\bibfield  {journal}
		{\bibinfo  {journal} {Phys. Rev. X}\ }\textbf {\bibinfo {volume} {11}},\
		\bibinfo {pages} {021053} (\bibinfo {year} {2021}{\natexlab{a}})},\ \Eprint
	{http://arxiv.org/abs/2010.14527} {arXiv:2010.14527 [gr-qc]} \BibitemShut
	{NoStop}%
	\bibitem [{\citenamefont {Abbott}\ \emph {et~al.}(2024)\citenamefont {Abbott}
		\emph {et~al.}}]{LIGOScientific:2021usb}%
	\BibitemOpen
	\bibfield  {author} {\bibinfo {author} {\bibfnamefont {R.}~\bibnamefont
			{Abbott}} \emph {et~al.} (\bibinfo {collaboration} {LIGO Scientific,
			VIRGO}),\ }\href {\doibase 10.1103/PhysRevD.109.022001} {\bibfield  {journal}
		{\bibinfo  {journal} {Phys. Rev. D}\ }\textbf {\bibinfo {volume} {109}},\
		\bibinfo {pages} {022001} (\bibinfo {year} {2024})},\ \Eprint
	{http://arxiv.org/abs/2108.01045} {arXiv:2108.01045 [gr-qc]} \BibitemShut
	{NoStop}%
	\bibitem [{\citenamefont {Abbott}\ \emph {et~al.}(2023)\citenamefont {Abbott}
		\emph {et~al.}}]{KAGRA:2021vkt}%
	\BibitemOpen
	\bibfield  {author} {\bibinfo {author} {\bibfnamefont {R.}~\bibnamefont
			{Abbott}} \emph {et~al.} (\bibinfo {collaboration} {KAGRA, VIRGO, LIGO
			Scientific}),\ }\href {\doibase 10.1103/PhysRevX.13.041039} {\bibfield
		{journal} {\bibinfo  {journal} {Phys. Rev. X}\ }\textbf {\bibinfo {volume}
			{13}},\ \bibinfo {pages} {041039} (\bibinfo {year} {2023})},\ \Eprint
	{http://arxiv.org/abs/2111.03606} {arXiv:2111.03606 [gr-qc]} \BibitemShut
	{NoStop}%
	\bibitem [{\citenamefont {Abbott}\ \emph
		{et~al.}(2020{\natexlab{a}})\citenamefont {Abbott} \emph
		{et~al.}}]{LIGOScientific:2019hgc}%
	\BibitemOpen
	\bibfield  {author} {\bibinfo {author} {\bibfnamefont {B.~P.}\ \bibnamefont
			{Abbott}} \emph {et~al.} (\bibinfo {collaboration} {LIGO Scientific,
			Virgo}),\ }\href {\doibase 10.1088/1361-6382/ab685e} {\bibfield  {journal}
		{\bibinfo  {journal} {Class. Quant. Grav.}\ }\textbf {\bibinfo {volume}
			{37}},\ \bibinfo {pages} {055002} (\bibinfo {year} {2020}{\natexlab{a}})},\
	\Eprint {http://arxiv.org/abs/1908.11170} {arXiv:1908.11170 [gr-qc]}
	\BibitemShut {NoStop}%
	\bibitem [{\citenamefont {Davis}\ \emph {et~al.}(2021)\citenamefont {Davis}
		\emph {et~al.}}]{LIGO:2021ppb}%
	\BibitemOpen
	\bibfield  {author} {\bibinfo {author} {\bibfnamefont {D.}~\bibnamefont
			{Davis}} \emph {et~al.} (\bibinfo {collaboration} {LIGO}),\ }\href {\doibase
		10.1088/1361-6382/abfd85} {\bibfield  {journal} {\bibinfo  {journal} {Class.
				Quant. Grav.}\ }\textbf {\bibinfo {volume} {38}},\ \bibinfo {pages} {135014}
		(\bibinfo {year} {2021})},\ \Eprint {http://arxiv.org/abs/2101.11673}
	{arXiv:2101.11673 [astro-ph.IM]} \BibitemShut {NoStop}%
	\bibitem [{\citenamefont {Soni}\ \emph {et~al.}(2025)\citenamefont {Soni} \emph
		{et~al.}}]{LIGO:2024kkz}%
	\BibitemOpen
	\bibfield  {author} {\bibinfo {author} {\bibfnamefont {S.}~\bibnamefont
			{Soni}} \emph {et~al.} (\bibinfo {collaboration} {LIGO}),\ }\href {\doibase
		10.1088/1361-6382/adc4b6} {\bibfield  {journal} {\bibinfo  {journal} {Class.
				Quant. Grav.}\ }\textbf {\bibinfo {volume} {42}},\ \bibinfo {pages} {085016}
		(\bibinfo {year} {2025})},\ \Eprint {http://arxiv.org/abs/2409.02831}
	{arXiv:2409.02831 [astro-ph.IM]} \BibitemShut {NoStop}%
	\bibitem [{\citenamefont {Powell}(2018)}]{Powell:2018csz}%
	\BibitemOpen
	\bibfield  {author} {\bibinfo {author} {\bibfnamefont {J.}~\bibnamefont
			{Powell}},\ }\href {\doibase 10.1088/1361-6382/aacf18} {\bibfield  {journal}
		{\bibinfo  {journal} {Class. Quant. Grav.}\ }\textbf {\bibinfo {volume}
			{35}},\ \bibinfo {pages} {155017} (\bibinfo {year} {2018})},\ \Eprint
	{http://arxiv.org/abs/1803.11346} {arXiv:1803.11346 [astro-ph.IM]}
	\BibitemShut {NoStop}%
	\bibitem [{\citenamefont {Pankow}\ \emph {et~al.}(2018)\citenamefont {Pankow}
		\emph {et~al.}}]{Pankow:2018qpo}%
	\BibitemOpen
	\bibfield  {author} {\bibinfo {author} {\bibfnamefont {C.}~\bibnamefont
			{Pankow}} \emph {et~al.},\ }\href {\doibase 10.1103/PhysRevD.98.084016}
	{\bibfield  {journal} {\bibinfo  {journal} {Phys. Rev. D}\ }\textbf {\bibinfo
			{volume} {98}},\ \bibinfo {pages} {084016} (\bibinfo {year} {2018})},\
	\Eprint {http://arxiv.org/abs/1808.03619} {arXiv:1808.03619 [gr-qc]}
	\BibitemShut {NoStop}%
	\bibitem [{\citenamefont {Cornish}\ \emph {et~al.}(2021)\citenamefont
		{Cornish}, \citenamefont {Littenberg}, \citenamefont {B{\'e}csy},
		\citenamefont {Chatziioannou}, \citenamefont {Clark}, \citenamefont
		{Ghonge},\ and\ \citenamefont {Millhouse}}]{Cornish:2020dwh}%
	\BibitemOpen
	\bibfield  {author} {\bibinfo {author} {\bibfnamefont {N.~J.}\ \bibnamefont
			{Cornish}}, \bibinfo {author} {\bibfnamefont {T.~B.}\ \bibnamefont
			{Littenberg}}, \bibinfo {author} {\bibfnamefont {B.}~\bibnamefont
			{B{\'e}csy}}, \bibinfo {author} {\bibfnamefont {K.}~\bibnamefont
			{Chatziioannou}}, \bibinfo {author} {\bibfnamefont {J.~A.}\ \bibnamefont
			{Clark}}, \bibinfo {author} {\bibfnamefont {S.}~\bibnamefont {Ghonge}}, \
		and\ \bibinfo {author} {\bibfnamefont {M.}~\bibnamefont {Millhouse}},\ }\href
	{\doibase 10.1103/PhysRevD.103.044006} {\bibfield  {journal} {\bibinfo
			{journal} {Phys. Rev. D}\ }\textbf {\bibinfo {volume} {103}},\ \bibinfo
		{pages} {044006} (\bibinfo {year} {2021})},\ \Eprint
	{http://arxiv.org/abs/2011.09494} {arXiv:2011.09494 [gr-qc]} \BibitemShut
	{NoStop}%
	\bibitem [{\citenamefont {Merritt}\ \emph {et~al.}(2021)\citenamefont
		{Merritt}, \citenamefont {Farr}, \citenamefont {Hur}, \citenamefont
		{Edelman},\ and\ \citenamefont {Doctor}}]{Merritt:2021xwh}%
	\BibitemOpen
	\bibfield  {author} {\bibinfo {author} {\bibfnamefont {J.}~\bibnamefont
			{Merritt}}, \bibinfo {author} {\bibfnamefont {B.}~\bibnamefont {Farr}},
		\bibinfo {author} {\bibfnamefont {R.}~\bibnamefont {Hur}}, \bibinfo {author}
		{\bibfnamefont {B.}~\bibnamefont {Edelman}}, \ and\ \bibinfo {author}
		{\bibfnamefont {Z.}~\bibnamefont {Doctor}},\ }\href {\doibase
		10.1103/PhysRevD.104.102004} {\bibfield  {journal} {\bibinfo  {journal}
			{Phys. Rev. D}\ }\textbf {\bibinfo {volume} {104}},\ \bibinfo {pages}
		{102004} (\bibinfo {year} {2021})},\ \Eprint
	{http://arxiv.org/abs/2108.12044} {arXiv:2108.12044 [gr-qc]} \BibitemShut
	{NoStop}%
	\bibitem [{\citenamefont {Kwok}\ \emph {et~al.}(2022)\citenamefont {Kwok},
		\citenamefont {Lo}, \citenamefont {Weinstein},\ and\ \citenamefont
		{Li}}]{Kwok:2021zny}%
	\BibitemOpen
	\bibfield  {author} {\bibinfo {author} {\bibfnamefont {J.~Y.~L.}\
			\bibnamefont {Kwok}}, \bibinfo {author} {\bibfnamefont {R.~K.~L.}\
			\bibnamefont {Lo}}, \bibinfo {author} {\bibfnamefont {A.~J.}\ \bibnamefont
			{Weinstein}}, \ and\ \bibinfo {author} {\bibfnamefont {T.~G.~F.}\
			\bibnamefont {Li}},\ }\href {\doibase 10.1103/PhysRevD.105.024066} {\bibfield
		{journal} {\bibinfo  {journal} {Phys. Rev. D}\ }\textbf {\bibinfo {volume}
			{105}},\ \bibinfo {pages} {024066} (\bibinfo {year} {2022})},\ \Eprint
	{http://arxiv.org/abs/2109.07642} {arXiv:2109.07642 [gr-qc]} \BibitemShut
	{NoStop}%
	\bibitem [{\citenamefont {Hourihane}\ \emph {et~al.}(2022)\citenamefont
		{Hourihane}, \citenamefont {Chatziioannou}, \citenamefont {Wijngaarden},
		\citenamefont {Davis}, \citenamefont {Littenberg},\ and\ \citenamefont
		{Cornish}}]{Hourihane:2022doe}%
	\BibitemOpen
	\bibfield  {author} {\bibinfo {author} {\bibfnamefont {S.}~\bibnamefont
			{Hourihane}}, \bibinfo {author} {\bibfnamefont {K.}~\bibnamefont
			{Chatziioannou}}, \bibinfo {author} {\bibfnamefont {M.}~\bibnamefont
			{Wijngaarden}}, \bibinfo {author} {\bibfnamefont {D.}~\bibnamefont {Davis}},
		\bibinfo {author} {\bibfnamefont {T.}~\bibnamefont {Littenberg}}, \ and\
		\bibinfo {author} {\bibfnamefont {N.}~\bibnamefont {Cornish}},\ }\href
	{\doibase 10.1103/PhysRevD.106.042006} {\bibfield  {journal} {\bibinfo
			{journal} {Phys. Rev. D}\ }\textbf {\bibinfo {volume} {106}},\ \bibinfo
		{pages} {042006} (\bibinfo {year} {2022})},\ \Eprint
	{http://arxiv.org/abs/2205.13580} {arXiv:2205.13580 [gr-qc]} \BibitemShut
	{NoStop}%
	\bibitem [{\citenamefont {Davis}\ \emph {et~al.}(2022)\citenamefont {Davis},
		\citenamefont {Littenberg}, \citenamefont {Romero-Shaw}, \citenamefont
		{Millhouse}, \citenamefont {McIver}, \citenamefont {Di~Renzo},\ and\
		\citenamefont {Ashton}}]{Davis:2022ird}%
	\BibitemOpen
	\bibfield  {author} {\bibinfo {author} {\bibfnamefont {D.}~\bibnamefont
			{Davis}}, \bibinfo {author} {\bibfnamefont {T.~B.}\ \bibnamefont
			{Littenberg}}, \bibinfo {author} {\bibfnamefont {I.~M.}\ \bibnamefont
			{Romero-Shaw}}, \bibinfo {author} {\bibfnamefont {M.}~\bibnamefont
			{Millhouse}}, \bibinfo {author} {\bibfnamefont {J.}~\bibnamefont {McIver}},
		\bibinfo {author} {\bibfnamefont {F.}~\bibnamefont {Di~Renzo}}, \ and\
		\bibinfo {author} {\bibfnamefont {G.}~\bibnamefont {Ashton}},\ }\href
	{\doibase 10.1088/1361-6382/aca238} {\bibfield  {journal} {\bibinfo
			{journal} {Class. Quant. Grav.}\ }\textbf {\bibinfo {volume} {39}},\ \bibinfo
		{pages} {245013} (\bibinfo {year} {2022})},\ \Eprint
	{http://arxiv.org/abs/2207.03429} {arXiv:2207.03429 [astro-ph.IM]}
	\BibitemShut {NoStop}%
	\bibitem [{\citenamefont {Mohanty}\ and\ \citenamefont
		{Chowdhury}(2023)}]{Mohanty:2023mjn}%
	\BibitemOpen
	\bibfield  {author} {\bibinfo {author} {\bibfnamefont {S.~D.}\ \bibnamefont
			{Mohanty}}\ and\ \bibinfo {author} {\bibfnamefont {M.~A.~T.}\ \bibnamefont
			{Chowdhury}},\ }\href {\doibase 10.1088/1361-6382/acd0fe} {\bibfield
		{journal} {\bibinfo  {journal} {Class. Quant. Grav.}\ }\textbf {\bibinfo
			{volume} {40}},\ \bibinfo {pages} {125001} (\bibinfo {year} {2023})},\
	\Eprint {http://arxiv.org/abs/2301.02398} {arXiv:2301.02398 [gr-qc]}
	\BibitemShut {NoStop}%
	\bibitem [{\citenamefont {Narola}\ \emph {et~al.}(2025)\citenamefont {Narola}
		\emph {et~al.}}]{Narola:2024qdh}%
	\BibitemOpen
	\bibfield  {author} {\bibinfo {author} {\bibfnamefont {H.}~\bibnamefont
			{Narola}} \emph {et~al.},\ }\href {\doibase 10.1103/l6tp-ykxp} {\bibfield
		{journal} {\bibinfo  {journal} {Phys. Rev. D}\ }\textbf {\bibinfo {volume}
			{112}},\ \bibinfo {pages} {024079} (\bibinfo {year} {2025})},\ \Eprint
	{http://arxiv.org/abs/2411.15506} {arXiv:2411.15506 [gr-qc]} \BibitemShut
	{NoStop}%
	\bibitem [{\citenamefont {Hinderer}\ and\ \citenamefont
		{Babak}(2017)}]{Hinderer:2017jcs}%
	\BibitemOpen
	\bibfield  {author} {\bibinfo {author} {\bibfnamefont {T.}~\bibnamefont
			{Hinderer}}\ and\ \bibinfo {author} {\bibfnamefont {S.}~\bibnamefont
			{Babak}},\ }\href {\doibase 10.1103/PhysRevD.96.104048} {\bibfield  {journal}
		{\bibinfo  {journal} {Phys. Rev. D}\ }\textbf {\bibinfo {volume} {96}},\
		\bibinfo {pages} {104048} (\bibinfo {year} {2017})},\ \Eprint
	{http://arxiv.org/abs/1707.08426} {arXiv:1707.08426 [gr-qc]} \BibitemShut
	{NoStop}%
	\bibitem [{\citenamefont {Cao}\ and\ \citenamefont {Han}(2017)}]{Cao:2017ndf}%
	\BibitemOpen
	\bibfield  {author} {\bibinfo {author} {\bibfnamefont {Z.}~\bibnamefont
			{Cao}}\ and\ \bibinfo {author} {\bibfnamefont {W.-B.}\ \bibnamefont {Han}},\
	}\href {\doibase 10.1103/PhysRevD.96.044028} {\bibfield  {journal} {\bibinfo
			{journal} {Phys. Rev. D}\ }\textbf {\bibinfo {volume} {96}},\ \bibinfo
		{pages} {044028} (\bibinfo {year} {2017})},\ \Eprint
	{http://arxiv.org/abs/1708.00166} {arXiv:1708.00166 [gr-qc]} \BibitemShut
	{NoStop}%
	\bibitem [{\citenamefont {Liu}\ \emph {et~al.}(2022)\citenamefont {Liu},
		\citenamefont {Cao},\ and\ \citenamefont {Zhu}}]{Liu:2021pkr}%
	\BibitemOpen
	\bibfield  {author} {\bibinfo {author} {\bibfnamefont {X.}~\bibnamefont
			{Liu}}, \bibinfo {author} {\bibfnamefont {Z.}~\bibnamefont {Cao}}, \ and\
		\bibinfo {author} {\bibfnamefont {Z.-H.}\ \bibnamefont {Zhu}},\ }\href
	{\doibase 10.1088/1361-6382/ac4119} {\bibfield  {journal} {\bibinfo
			{journal} {Class. Quant. Grav.}\ }\textbf {\bibinfo {volume} {39}},\ \bibinfo
		{pages} {035009} (\bibinfo {year} {2022})},\ \Eprint
	{http://arxiv.org/abs/2102.08614} {arXiv:2102.08614 [gr-qc]} \BibitemShut
	{NoStop}%
	\bibitem [{\citenamefont {Nagar}\ \emph {et~al.}(2021)\citenamefont {Nagar},
		\citenamefont {Bonino},\ and\ \citenamefont {Rettegno}}]{Nagar:2021gss}%
	\BibitemOpen
	\bibfield  {author} {\bibinfo {author} {\bibfnamefont {A.}~\bibnamefont
			{Nagar}}, \bibinfo {author} {\bibfnamefont {A.}~\bibnamefont {Bonino}}, \
		and\ \bibinfo {author} {\bibfnamefont {P.}~\bibnamefont {Rettegno}},\ }\href
	{\doibase 10.1103/PhysRevD.103.104021} {\bibfield  {journal} {\bibinfo
			{journal} {Phys. Rev. D}\ }\textbf {\bibinfo {volume} {103}},\ \bibinfo
		{pages} {104021} (\bibinfo {year} {2021})},\ \Eprint
	{http://arxiv.org/abs/2101.08624} {arXiv:2101.08624 [gr-qc]} \BibitemShut
	{NoStop}%
	\bibitem [{\citenamefont {Khalil}\ \emph {et~al.}(2021)\citenamefont {Khalil},
		\citenamefont {Buonanno}, \citenamefont {Steinhoff},\ and\ \citenamefont
		{Vines}}]{Khalil:2021txt}%
	\BibitemOpen
	\bibfield  {author} {\bibinfo {author} {\bibfnamefont {M.}~\bibnamefont
			{Khalil}}, \bibinfo {author} {\bibfnamefont {A.}~\bibnamefont {Buonanno}},
		\bibinfo {author} {\bibfnamefont {J.}~\bibnamefont {Steinhoff}}, \ and\
		\bibinfo {author} {\bibfnamefont {J.}~\bibnamefont {Vines}},\ }\href
	{\doibase 10.1103/PhysRevD.104.024046} {\bibfield  {journal} {\bibinfo
			{journal} {Phys. Rev. D}\ }\textbf {\bibinfo {volume} {104}},\ \bibinfo
		{pages} {024046} (\bibinfo {year} {2021})},\ \Eprint
	{http://arxiv.org/abs/2104.11705} {arXiv:2104.11705 [gr-qc]} \BibitemShut
	{NoStop}%
	\bibitem [{\citenamefont {Islam}\ \emph {et~al.}(2021)\citenamefont {Islam},
		\citenamefont {Varma}, \citenamefont {Lodman}, \citenamefont {Field},
		\citenamefont {Khanna}, \citenamefont {Scheel}, \citenamefont {Pfeiffer},
		\citenamefont {Gerosa},\ and\ \citenamefont {Kidder}}]{Islam:2021mha}%
	\BibitemOpen
	\bibfield  {author} {\bibinfo {author} {\bibfnamefont {T.}~\bibnamefont
			{Islam}}, \bibinfo {author} {\bibfnamefont {V.}~\bibnamefont {Varma}},
		\bibinfo {author} {\bibfnamefont {J.}~\bibnamefont {Lodman}}, \bibinfo
		{author} {\bibfnamefont {S.~E.}\ \bibnamefont {Field}}, \bibinfo {author}
		{\bibfnamefont {G.}~\bibnamefont {Khanna}}, \bibinfo {author} {\bibfnamefont
			{M.~A.}\ \bibnamefont {Scheel}}, \bibinfo {author} {\bibfnamefont {H.~P.}\
			\bibnamefont {Pfeiffer}}, \bibinfo {author} {\bibfnamefont {D.}~\bibnamefont
			{Gerosa}}, \ and\ \bibinfo {author} {\bibfnamefont {L.~E.}\ \bibnamefont
			{Kidder}},\ }\href {\doibase 10.1103/PhysRevD.103.064022} {\bibfield
		{journal} {\bibinfo  {journal} {Phys. Rev. D}\ }\textbf {\bibinfo {volume}
			{103}},\ \bibinfo {pages} {064022} (\bibinfo {year} {2021})},\ \Eprint
	{http://arxiv.org/abs/2101.11798} {arXiv:2101.11798 [gr-qc]} \BibitemShut
	{NoStop}%
	\bibitem [{\citenamefont {Ramos-Buades}\ \emph {et~al.}(2022)\citenamefont
		{Ramos-Buades}, \citenamefont {Buonanno}, \citenamefont {Khalil},\ and\
		\citenamefont {Ossokine}}]{Ramos-Buades:2021adz}%
	\BibitemOpen
	\bibfield  {author} {\bibinfo {author} {\bibfnamefont {A.}~\bibnamefont
			{Ramos-Buades}}, \bibinfo {author} {\bibfnamefont {A.}~\bibnamefont
			{Buonanno}}, \bibinfo {author} {\bibfnamefont {M.}~\bibnamefont {Khalil}}, \
		and\ \bibinfo {author} {\bibfnamefont {S.}~\bibnamefont {Ossokine}},\ }\href
	{\doibase 10.1103/PhysRevD.105.044035} {\bibfield  {journal} {\bibinfo
			{journal} {Phys. Rev. D}\ }\textbf {\bibinfo {volume} {105}},\ \bibinfo
		{pages} {044035} (\bibinfo {year} {2022})},\ \Eprint
	{http://arxiv.org/abs/2112.06952} {arXiv:2112.06952 [gr-qc]} \BibitemShut
	{NoStop}%
	\bibitem [{\citenamefont {Wang}\ \emph {et~al.}(2023)\citenamefont {Wang},
		\citenamefont {Zou},\ and\ \citenamefont {Liu}}]{Wang:2023ueg}%
	\BibitemOpen
	\bibfield  {author} {\bibinfo {author} {\bibfnamefont {H.}~\bibnamefont
			{Wang}}, \bibinfo {author} {\bibfnamefont {Y.-C.}\ \bibnamefont {Zou}}, \
		and\ \bibinfo {author} {\bibfnamefont {Y.}~\bibnamefont {Liu}},\ }\href
	{\doibase 10.1103/PhysRevD.107.124061} {\bibfield  {journal} {\bibinfo
			{journal} {Phys. Rev. D}\ }\textbf {\bibinfo {volume} {107}},\ \bibinfo
		{pages} {124061} (\bibinfo {year} {2023})},\ \Eprint
	{http://arxiv.org/abs/2302.11227} {arXiv:2302.11227 [gr-qc]} \BibitemShut
	{NoStop}%
	\bibitem [{\citenamefont {Abac}\ \emph {et~al.}(2024)\citenamefont {Abac} \emph
		{et~al.}}]{LIGOScientific:2023lpe}%
	\BibitemOpen
	\bibfield  {author} {\bibinfo {author} {\bibfnamefont {A.~G.}\ \bibnamefont
			{Abac}} \emph {et~al.} (\bibinfo {collaboration} {LIGO Scientific, KAGRA,
			VIRGO}),\ }\href {\doibase 10.3847/1538-4357/ad65ce} {\bibfield  {journal}
		{\bibinfo  {journal} {Astrophys. J.}\ }\textbf {\bibinfo {volume} {973}},\
		\bibinfo {pages} {132} (\bibinfo {year} {2024})},\ \Eprint
	{http://arxiv.org/abs/2308.03822} {arXiv:2308.03822 [astro-ph.HE]}
	\BibitemShut {NoStop}%
	\bibitem [{\citenamefont {Gupte}\ \emph {et~al.}(2025)\citenamefont {Gupte}
		\emph {et~al.}}]{Gupte:2024jfe}%
	\BibitemOpen
	\bibfield  {author} {\bibinfo {author} {\bibfnamefont {N.}~\bibnamefont
			{Gupte}} \emph {et~al.},\ }\href {\doibase 10.1103/vpyp-nvfp} {\bibfield
		{journal} {\bibinfo  {journal} {Phys. Rev. D}\ }\textbf {\bibinfo {volume}
			{112}},\ \bibinfo {pages} {104045} (\bibinfo {year} {2025})},\ \Eprint
	{http://arxiv.org/abs/2404.14286} {arXiv:2404.14286 [gr-qc]} \BibitemShut
	{NoStop}%
	\bibitem [{\citenamefont {Trenado}\ \emph {et~al.}(2025)\citenamefont
		{Trenado}, \citenamefont {Andrade}, \citenamefont {Climent},\ and\
		\citenamefont {Ferrer}}]{Trenado:2025ccf}%
	\BibitemOpen
	\bibfield  {author} {\bibinfo {author} {\bibfnamefont {J.}~\bibnamefont
			{Trenado}}, \bibinfo {author} {\bibfnamefont {T.}~\bibnamefont {Andrade}},
		\bibinfo {author} {\bibfnamefont {A.}~\bibnamefont {Climent}}, \ and\
		\bibinfo {author} {\bibfnamefont {M.~A.}\ \bibnamefont {Ferrer}},\
	}\href@noop {} {\  (\bibinfo {year} {2025})},\ \Eprint
	{http://arxiv.org/abs/2509.05269} {arXiv:2509.05269 [gr-qc]} \BibitemShut
	{NoStop}%
	\bibitem [{\citenamefont {Barausse}\ and\ \citenamefont
		{Rezzolla}(2008)}]{Barausse:2007dy}%
	\BibitemOpen
	\bibfield  {author} {\bibinfo {author} {\bibfnamefont {E.}~\bibnamefont
			{Barausse}}\ and\ \bibinfo {author} {\bibfnamefont {L.}~\bibnamefont
			{Rezzolla}},\ }\href {\doibase 10.1103/PhysRevD.77.104027} {\bibfield
		{journal} {\bibinfo  {journal} {Phys. Rev. D}\ }\textbf {\bibinfo {volume}
			{77}},\ \bibinfo {pages} {104027} (\bibinfo {year} {2008})},\ \Eprint
	{http://arxiv.org/abs/0711.4558} {arXiv:0711.4558 [gr-qc]} \BibitemShut
	{NoStop}%
	\bibitem [{\citenamefont {Dai}\ \emph {et~al.}(2022)\citenamefont {Dai},
		\citenamefont {Gong}, \citenamefont {Jiang},\ and\ \citenamefont
		{Liang}}]{Dai:2021olt}%
	\BibitemOpen
	\bibfield  {author} {\bibinfo {author} {\bibfnamefont {N.}~\bibnamefont
			{Dai}}, \bibinfo {author} {\bibfnamefont {Y.}~\bibnamefont {Gong}}, \bibinfo
		{author} {\bibfnamefont {T.}~\bibnamefont {Jiang}}, \ and\ \bibinfo {author}
		{\bibfnamefont {D.}~\bibnamefont {Liang}},\ }\href {\doibase
		10.1103/PhysRevD.106.064003} {\bibfield  {journal} {\bibinfo  {journal}
			{Phys. Rev. D}\ }\textbf {\bibinfo {volume} {106}},\ \bibinfo {pages}
		{064003} (\bibinfo {year} {2022})},\ \Eprint
	{http://arxiv.org/abs/2111.13514} {arXiv:2111.13514 [gr-qc]} \BibitemShut
	{NoStop}%
	\bibitem [{\citenamefont {Cole}\ \emph {et~al.}(2023)\citenamefont {Cole},
		\citenamefont {Coogan}, \citenamefont {Kavanagh},\ and\ \citenamefont
		{Bertone}}]{Cole:2022ucw}%
	\BibitemOpen
	\bibfield  {author} {\bibinfo {author} {\bibfnamefont {P.~S.}\ \bibnamefont
			{Cole}}, \bibinfo {author} {\bibfnamefont {A.}~\bibnamefont {Coogan}},
		\bibinfo {author} {\bibfnamefont {B.~J.}\ \bibnamefont {Kavanagh}}, \ and\
		\bibinfo {author} {\bibfnamefont {G.}~\bibnamefont {Bertone}},\ }\href
	{\doibase 10.1103/PhysRevD.107.083006} {\bibfield  {journal} {\bibinfo
			{journal} {Phys. Rev. D}\ }\textbf {\bibinfo {volume} {107}},\ \bibinfo
		{pages} {083006} (\bibinfo {year} {2023})},\ \Eprint
	{http://arxiv.org/abs/2207.07576} {arXiv:2207.07576 [astro-ph.CO]}
	\BibitemShut {NoStop}%
	\bibitem [{\citenamefont {Garg}\ \emph {et~al.}(2022)\citenamefont {Garg},
		\citenamefont {Derdzinski}, \citenamefont {Zwick}, \citenamefont {Capelo},\
		and\ \citenamefont {Mayer}}]{Garg:2022nko}%
	\BibitemOpen
	\bibfield  {author} {\bibinfo {author} {\bibfnamefont {M.}~\bibnamefont
			{Garg}}, \bibinfo {author} {\bibfnamefont {A.}~\bibnamefont {Derdzinski}},
		\bibinfo {author} {\bibfnamefont {L.}~\bibnamefont {Zwick}}, \bibinfo
		{author} {\bibfnamefont {P.~R.}\ \bibnamefont {Capelo}}, \ and\ \bibinfo
		{author} {\bibfnamefont {L.}~\bibnamefont {Mayer}},\ }\href {\doibase
		10.1093/mnras/stac2711} {\bibfield  {journal} {\bibinfo  {journal} {Mon. Not.
				Roy. Astron. Soc.}\ }\textbf {\bibinfo {volume} {517}},\ \bibinfo {pages}
		{1339} (\bibinfo {year} {2022})},\ \Eprint {http://arxiv.org/abs/2206.05292}
	{arXiv:2206.05292 [astro-ph.GA]} \BibitemShut {NoStop}%
	\bibitem [{\citenamefont {Dai}\ \emph {et~al.}(2024)\citenamefont {Dai},
		\citenamefont {Gong}, \citenamefont {Zhao},\ and\ \citenamefont
		{Jiang}}]{Dai:2023cft}%
	\BibitemOpen
	\bibfield  {author} {\bibinfo {author} {\bibfnamefont {N.}~\bibnamefont
			{Dai}}, \bibinfo {author} {\bibfnamefont {Y.}~\bibnamefont {Gong}}, \bibinfo
		{author} {\bibfnamefont {Y.}~\bibnamefont {Zhao}}, \ and\ \bibinfo {author}
		{\bibfnamefont {T.}~\bibnamefont {Jiang}},\ }\href {\doibase
		10.1103/PhysRevD.110.084080} {\bibfield  {journal} {\bibinfo  {journal}
			{Phys. Rev. D}\ }\textbf {\bibinfo {volume} {110}},\ \bibinfo {pages}
		{084080} (\bibinfo {year} {2024})},\ \Eprint
	{http://arxiv.org/abs/2301.05088} {arXiv:2301.05088 [gr-qc]} \BibitemShut
	{NoStop}%
	\bibitem [{\citenamefont {Zwick}\ \emph {et~al.}(2024)\citenamefont {Zwick},
		\citenamefont {Tiede}, \citenamefont {Trani}, \citenamefont {Derdzinski},
		\citenamefont {Haiman}, \citenamefont {D'Orazio},\ and\ \citenamefont
		{Samsing}}]{Zwick:2024yzh}%
	\BibitemOpen
	\bibfield  {author} {\bibinfo {author} {\bibfnamefont {L.}~\bibnamefont
			{Zwick}}, \bibinfo {author} {\bibfnamefont {C.}~\bibnamefont {Tiede}},
		\bibinfo {author} {\bibfnamefont {A.~A.}\ \bibnamefont {Trani}}, \bibinfo
		{author} {\bibfnamefont {A.}~\bibnamefont {Derdzinski}}, \bibinfo {author}
		{\bibfnamefont {Z.}~\bibnamefont {Haiman}}, \bibinfo {author} {\bibfnamefont
			{D.~J.}\ \bibnamefont {D'Orazio}}, \ and\ \bibinfo {author} {\bibfnamefont
			{J.}~\bibnamefont {Samsing}},\ }\href {\doibase 10.1103/PhysRevD.110.103005}
	{\bibfield  {journal} {\bibinfo  {journal} {Phys. Rev. D}\ }\textbf {\bibinfo
			{volume} {110}},\ \bibinfo {pages} {103005} (\bibinfo {year} {2024})},\
	\Eprint {http://arxiv.org/abs/2405.05698} {arXiv:2405.05698 [gr-qc]}
	\BibitemShut {NoStop}%
	\bibitem [{\citenamefont {Zwick}\ \emph {et~al.}(2025)\citenamefont {Zwick},
		\citenamefont {Tak{\'a}tsy}, \citenamefont {Saini}, \citenamefont {Hendriks},
		\citenamefont {Samsing}, \citenamefont {Tiede}, \citenamefont {Rowan},\ and\
		\citenamefont {Trani}}]{Zwick:2025wkt}%
	\BibitemOpen
	\bibfield  {author} {\bibinfo {author} {\bibfnamefont {L.}~\bibnamefont
			{Zwick}}, \bibinfo {author} {\bibfnamefont {J.}~\bibnamefont {Tak{\'a}tsy}},
		\bibinfo {author} {\bibfnamefont {P.}~\bibnamefont {Saini}}, \bibinfo
		{author} {\bibfnamefont {K.}~\bibnamefont {Hendriks}}, \bibinfo {author}
		{\bibfnamefont {J.}~\bibnamefont {Samsing}}, \bibinfo {author} {\bibfnamefont
			{C.}~\bibnamefont {Tiede}}, \bibinfo {author} {\bibfnamefont
			{C.}~\bibnamefont {Rowan}}, \ and\ \bibinfo {author} {\bibfnamefont {A.~A.}\
			\bibnamefont {Trani}},\ }\href {\doibase 10.3847/1538-4357/adf6b8} {\bibfield
		{journal} {\bibinfo  {journal} {Astrophys. J.}\ }\textbf {\bibinfo {volume}
			{991}},\ \bibinfo {pages} {131} (\bibinfo {year} {2025})},\ \Eprint
	{http://arxiv.org/abs/2503.24084} {arXiv:2503.24084 [astro-ph.HE]}
	\BibitemShut {NoStop}%
	\bibitem [{\citenamefont {Torigoe}\ \emph {et~al.}(2009)\citenamefont
		{Torigoe}, \citenamefont {Hattori},\ and\ \citenamefont
		{Asada}}]{Torigoe:2009bw}%
	\BibitemOpen
	\bibfield  {author} {\bibinfo {author} {\bibfnamefont {Y.}~\bibnamefont
			{Torigoe}}, \bibinfo {author} {\bibfnamefont {K.}~\bibnamefont {Hattori}}, \
		and\ \bibinfo {author} {\bibfnamefont {H.}~\bibnamefont {Asada}},\ }\href
	{\doibase 10.1103/PhysRevLett.102.251101} {\bibfield  {journal} {\bibinfo
			{journal} {Phys. Rev. Lett.}\ }\textbf {\bibinfo {volume} {102}},\ \bibinfo
		{pages} {251101} (\bibinfo {year} {2009})},\ \Eprint
	{http://arxiv.org/abs/0906.1448} {arXiv:0906.1448 [gr-qc]} \BibitemShut
	{NoStop}%
	\bibitem [{\citenamefont {Dmitra{\v{s}}inovi{\'c}}\ \emph
		{et~al.}(2014)\citenamefont {Dmitra{\v{s}}inovi{\'c}}, \citenamefont
		{{\v{S}}uvakov},\ and\ \citenamefont {Hudomal}}]{Dmitrasinovic:2014lha}%
	\BibitemOpen
	\bibfield  {author} {\bibinfo {author} {\bibfnamefont {V.}~\bibnamefont
			{Dmitra{\v{s}}inovi{\'c}}}, \bibinfo {author} {\bibfnamefont
			{M.}~\bibnamefont {{\v{S}}uvakov}}, \ and\ \bibinfo {author} {\bibfnamefont
			{A.}~\bibnamefont {Hudomal}},\ }\href {\doibase
		10.1103/PhysRevLett.113.101102} {\bibfield  {journal} {\bibinfo  {journal}
			{Phys. Rev. Lett.}\ }\textbf {\bibinfo {volume} {113}},\ \bibinfo {pages}
		{101102} (\bibinfo {year} {2014})},\ \Eprint
	{http://arxiv.org/abs/1501.03405} {arXiv:1501.03405 [gr-qc]} \BibitemShut
	{NoStop}%
	\bibitem [{\citenamefont {Bonetti}\ \emph {et~al.}(2017)\citenamefont
		{Bonetti}, \citenamefont {Barausse}, \citenamefont {Faye}, \citenamefont
		{Haardt},\ and\ \citenamefont {Sesana}}]{Bonetti:2017hnb}%
	\BibitemOpen
	\bibfield  {author} {\bibinfo {author} {\bibfnamefont {M.}~\bibnamefont
			{Bonetti}}, \bibinfo {author} {\bibfnamefont {E.}~\bibnamefont {Barausse}},
		\bibinfo {author} {\bibfnamefont {G.}~\bibnamefont {Faye}}, \bibinfo {author}
		{\bibfnamefont {F.}~\bibnamefont {Haardt}}, \ and\ \bibinfo {author}
		{\bibfnamefont {A.}~\bibnamefont {Sesana}},\ }\href {\doibase
		10.1088/1361-6382/aa8da5} {\bibfield  {journal} {\bibinfo  {journal} {Class.
				Quant. Grav.}\ }\textbf {\bibinfo {volume} {34}},\ \bibinfo {pages} {215004}
		(\bibinfo {year} {2017})},\ \Eprint {http://arxiv.org/abs/1707.04902}
	{arXiv:1707.04902 [gr-qc]} \BibitemShut {NoStop}%
	\bibitem [{\citenamefont {Gupta}\ \emph {et~al.}(2020)\citenamefont {Gupta},
		\citenamefont {Suzuki}, \citenamefont {Okawa},\ and\ \citenamefont
		{Maeda}}]{Gupta:2019unn}%
	\BibitemOpen
	\bibfield  {author} {\bibinfo {author} {\bibfnamefont {P.}~\bibnamefont
			{Gupta}}, \bibinfo {author} {\bibfnamefont {H.}~\bibnamefont {Suzuki}},
		\bibinfo {author} {\bibfnamefont {H.}~\bibnamefont {Okawa}}, \ and\ \bibinfo
		{author} {\bibfnamefont {K.-i.}\ \bibnamefont {Maeda}},\ }\href {\doibase
		10.1103/PhysRevD.101.104053} {\bibfield  {journal} {\bibinfo  {journal}
			{Phys. Rev. D}\ }\textbf {\bibinfo {volume} {101}},\ \bibinfo {pages}
		{104053} (\bibinfo {year} {2020})},\ \Eprint
	{http://arxiv.org/abs/1911.11318} {arXiv:1911.11318 [gr-qc]} \BibitemShut
	{NoStop}%
	\bibitem [{\citenamefont {Chandramouli}\ and\ \citenamefont
		{Yunes}(2022)}]{Chandramouli:2021kts}%
	\BibitemOpen
	\bibfield  {author} {\bibinfo {author} {\bibfnamefont {R.~S.}\ \bibnamefont
			{Chandramouli}}\ and\ \bibinfo {author} {\bibfnamefont {N.}~\bibnamefont
			{Yunes}},\ }\href {\doibase 10.1103/PhysRevD.105.064009} {\bibfield
		{journal} {\bibinfo  {journal} {Phys. Rev. D}\ }\textbf {\bibinfo {volume}
			{105}},\ \bibinfo {pages} {064009} (\bibinfo {year} {2022})},\ \Eprint
	{http://arxiv.org/abs/2107.00741} {arXiv:2107.00741 [gr-qc]} \BibitemShut
	{NoStop}%
	\bibitem [{\citenamefont {Samajdar}\ \emph {et~al.}(2021)\citenamefont
		{Samajdar}, \citenamefont {Janquart}, \citenamefont {Van Den~Broeck},\ and\
		\citenamefont {Dietrich}}]{Samajdar:2021egv}%
	\BibitemOpen
	\bibfield  {author} {\bibinfo {author} {\bibfnamefont {A.}~\bibnamefont
			{Samajdar}}, \bibinfo {author} {\bibfnamefont {J.}~\bibnamefont {Janquart}},
		\bibinfo {author} {\bibfnamefont {C.}~\bibnamefont {Van Den~Broeck}}, \ and\
		\bibinfo {author} {\bibfnamefont {T.}~\bibnamefont {Dietrich}},\ }\href
	{\doibase 10.1103/PhysRevD.104.044003} {\bibfield  {journal} {\bibinfo
			{journal} {Phys. Rev. D}\ }\textbf {\bibinfo {volume} {104}},\ \bibinfo
		{pages} {044003} (\bibinfo {year} {2021})},\ \Eprint
	{http://arxiv.org/abs/2102.07544} {arXiv:2102.07544 [gr-qc]} \BibitemShut
	{NoStop}%
	\bibitem [{\citenamefont {Relton}\ and\ \citenamefont
		{Raymond}(2021)}]{Relton:2021cax}%
	\BibitemOpen
	\bibfield  {author} {\bibinfo {author} {\bibfnamefont {P.}~\bibnamefont
			{Relton}}\ and\ \bibinfo {author} {\bibfnamefont {V.}~\bibnamefont
			{Raymond}},\ }\href {\doibase 10.1103/PhysRevD.104.084039} {\bibfield
		{journal} {\bibinfo  {journal} {Phys. Rev. D}\ }\textbf {\bibinfo {volume}
			{104}},\ \bibinfo {pages} {084039} (\bibinfo {year} {2021})},\ \Eprint
	{http://arxiv.org/abs/2103.16225} {arXiv:2103.16225 [gr-qc]} \BibitemShut
	{NoStop}%
	\bibitem [{\citenamefont {Antonelli}\ \emph {et~al.}(2021)\citenamefont
		{Antonelli}, \citenamefont {Burke},\ and\ \citenamefont
		{Gair}}]{Antonelli:2021vwg}%
	\BibitemOpen
	\bibfield  {author} {\bibinfo {author} {\bibfnamefont {A.}~\bibnamefont
			{Antonelli}}, \bibinfo {author} {\bibfnamefont {O.}~\bibnamefont {Burke}}, \
		and\ \bibinfo {author} {\bibfnamefont {J.~R.}\ \bibnamefont {Gair}},\ }\href
	{\doibase 10.1093/mnras/stab2358} {\bibfield  {journal} {\bibinfo  {journal}
			{Mon. Not. Roy. Astron. Soc.}\ }\textbf {\bibinfo {volume} {507}},\ \bibinfo
		{pages} {5069} (\bibinfo {year} {2021})},\ \Eprint
	{http://arxiv.org/abs/2104.01897} {arXiv:2104.01897 [gr-qc]} \BibitemShut
	{NoStop}%
	\bibitem [{\citenamefont {Wang}\ \emph {et~al.}(2024)\citenamefont {Wang},
		\citenamefont {Liang}, \citenamefont {Zhao}, \citenamefont {Liu},\ and\
		\citenamefont {Shao}}]{Wang:2023ldq}%
	\BibitemOpen
	\bibfield  {author} {\bibinfo {author} {\bibfnamefont {Z.}~\bibnamefont
			{Wang}}, \bibinfo {author} {\bibfnamefont {D.}~\bibnamefont {Liang}},
		\bibinfo {author} {\bibfnamefont {J.}~\bibnamefont {Zhao}}, \bibinfo {author}
		{\bibfnamefont {C.}~\bibnamefont {Liu}}, \ and\ \bibinfo {author}
		{\bibfnamefont {L.}~\bibnamefont {Shao}},\ }\href {\doibase
		10.1088/1361-6382/ad210b} {\bibfield  {journal} {\bibinfo  {journal} {Class.
				Quant. Grav.}\ }\textbf {\bibinfo {volume} {41}},\ \bibinfo {pages} {055011}
		(\bibinfo {year} {2024})},\ \Eprint {http://arxiv.org/abs/2304.06734}
	{arXiv:2304.06734 [astro-ph.IM]} \BibitemShut {NoStop}%
	\bibitem [{\citenamefont {Wang}\ \emph {et~al.}(2025)\citenamefont {Wang},
		\citenamefont {Liang},\ and\ \citenamefont {Shao}}]{Wang:2025aqk}%
	\BibitemOpen
	\bibfield  {author} {\bibinfo {author} {\bibfnamefont {Z.}~\bibnamefont
			{Wang}}, \bibinfo {author} {\bibfnamefont {D.}~\bibnamefont {Liang}}, \ and\
		\bibinfo {author} {\bibfnamefont {L.}~\bibnamefont {Shao}},\ }\href@noop {}
	{\  (\bibinfo {year} {2025})},\ \Eprint {http://arxiv.org/abs/2509.07737}
	{arXiv:2509.07737 [gr-qc]} \BibitemShut {NoStop}%
	\bibitem [{\citenamefont {Kosteleck{\'y}}\ and\ \citenamefont
		{Mewes}(2016)}]{Kostelecky:2016kfm}%
	\BibitemOpen
	\bibfield  {author} {\bibinfo {author} {\bibfnamefont {V.~A.}\ \bibnamefont
			{Kosteleck{\'y}}}\ and\ \bibinfo {author} {\bibfnamefont {M.}~\bibnamefont
			{Mewes}},\ }\href {\doibase 10.1016/j.physletb.2016.04.040} {\bibfield
		{journal} {\bibinfo  {journal} {Phys. Lett. B}\ }\textbf {\bibinfo {volume}
			{757}},\ \bibinfo {pages} {510} (\bibinfo {year} {2016})},\ \Eprint
	{http://arxiv.org/abs/1602.04782} {arXiv:1602.04782 [gr-qc]} \BibitemShut
	{NoStop}%
	\bibitem [{\citenamefont {Mewes}(2019)}]{Mewes:2019dhj}%
	\BibitemOpen
	\bibfield  {author} {\bibinfo {author} {\bibfnamefont {M.}~\bibnamefont
			{Mewes}},\ }\href {\doibase 10.1103/PhysRevD.99.104062} {\bibfield  {journal}
		{\bibinfo  {journal} {Phys. Rev. D}\ }\textbf {\bibinfo {volume} {99}},\
		\bibinfo {pages} {104062} (\bibinfo {year} {2019})},\ \Eprint
	{http://arxiv.org/abs/1905.00409} {arXiv:1905.00409 [gr-qc]} \BibitemShut
	{NoStop}%
	\bibitem [{\citenamefont {Shao}(2020)}]{Shao:2020shv}%
	\BibitemOpen
	\bibfield  {author} {\bibinfo {author} {\bibfnamefont {L.}~\bibnamefont
			{Shao}},\ }\href {\doibase 10.1103/PhysRevD.101.104019} {\bibfield  {journal}
		{\bibinfo  {journal} {Phys. Rev. D}\ }\textbf {\bibinfo {volume} {101}},\
		\bibinfo {pages} {104019} (\bibinfo {year} {2020})},\ \Eprint
	{http://arxiv.org/abs/2002.01185} {arXiv:2002.01185 [hep-ph]} \BibitemShut
	{NoStop}%
	\bibitem [{\citenamefont {Wang}\ \emph {et~al.}(2021)\citenamefont {Wang},
		\citenamefont {Shao},\ and\ \citenamefont {Liu}}]{Wang:2021ctl}%
	\BibitemOpen
	\bibfield  {author} {\bibinfo {author} {\bibfnamefont {Z.}~\bibnamefont
			{Wang}}, \bibinfo {author} {\bibfnamefont {L.}~\bibnamefont {Shao}}, \ and\
		\bibinfo {author} {\bibfnamefont {C.}~\bibnamefont {Liu}},\ }\href {\doibase
		10.3847/1538-4357/ac223c} {\bibfield  {journal} {\bibinfo  {journal}
			{Astrophys. J.}\ }\textbf {\bibinfo {volume} {921}},\ \bibinfo {pages} {158}
		(\bibinfo {year} {2021})},\ \Eprint {http://arxiv.org/abs/2108.02974}
	{arXiv:2108.02974 [gr-qc]} \BibitemShut {NoStop}%
	\bibitem [{\citenamefont {Zhao}\ \emph {et~al.}(2022)\citenamefont {Zhao},
		\citenamefont {Cao},\ and\ \citenamefont {Wang}}]{Zhao:2022pun}%
	\BibitemOpen
	\bibfield  {author} {\bibinfo {author} {\bibfnamefont {Z.-C.}\ \bibnamefont
			{Zhao}}, \bibinfo {author} {\bibfnamefont {Z.}~\bibnamefont {Cao}}, \ and\
		\bibinfo {author} {\bibfnamefont {S.}~\bibnamefont {Wang}},\ }\href {\doibase
		10.3847/1538-4357/ac62d3} {\bibfield  {journal} {\bibinfo  {journal}
			{Astrophys. J.}\ }\textbf {\bibinfo {volume} {930}},\ \bibinfo {pages} {139}
		(\bibinfo {year} {2022})},\ \Eprint {http://arxiv.org/abs/2201.02813}
	{arXiv:2201.02813 [gr-qc]} \BibitemShut {NoStop}%
	\bibitem [{\citenamefont {Haegel}\ \emph {et~al.}(2023)\citenamefont {Haegel},
		\citenamefont {O'Neal-Ault}, \citenamefont {Bailey}, \citenamefont {Tasson},
		\citenamefont {Bloom},\ and\ \citenamefont {Shao}}]{Haegel:2022ymk}%
	\BibitemOpen
	\bibfield  {author} {\bibinfo {author} {\bibfnamefont {L.}~\bibnamefont
			{Haegel}}, \bibinfo {author} {\bibfnamefont {K.}~\bibnamefont {O'Neal-Ault}},
		\bibinfo {author} {\bibfnamefont {Q.~G.}\ \bibnamefont {Bailey}}, \bibinfo
		{author} {\bibfnamefont {J.~D.}\ \bibnamefont {Tasson}}, \bibinfo {author}
		{\bibfnamefont {M.}~\bibnamefont {Bloom}}, \ and\ \bibinfo {author}
		{\bibfnamefont {L.}~\bibnamefont {Shao}},\ }\href {\doibase
		10.1103/PhysRevD.107.064031} {\bibfield  {journal} {\bibinfo  {journal}
			{Phys. Rev. D}\ }\textbf {\bibinfo {volume} {107}},\ \bibinfo {pages}
		{064031} (\bibinfo {year} {2023})},\ \Eprint
	{http://arxiv.org/abs/2210.04481} {arXiv:2210.04481 [gr-qc]} \BibitemShut
	{NoStop}%
	\bibitem [{\citenamefont {Zhu}\ \emph {et~al.}(2024)\citenamefont {Zhu},
		\citenamefont {Zhao}, \citenamefont {Yan}, \citenamefont {Wang},
		\citenamefont {Gong},\ and\ \citenamefont {Wang}}]{Zhu:2023rrx}%
	\BibitemOpen
	\bibfield  {author} {\bibinfo {author} {\bibfnamefont {T.}~\bibnamefont
			{Zhu}}, \bibinfo {author} {\bibfnamefont {W.}~\bibnamefont {Zhao}}, \bibinfo
		{author} {\bibfnamefont {J.-M.}\ \bibnamefont {Yan}}, \bibinfo {author}
		{\bibfnamefont {Y.-Z.}\ \bibnamefont {Wang}}, \bibinfo {author}
		{\bibfnamefont {C.}~\bibnamefont {Gong}}, \ and\ \bibinfo {author}
		{\bibfnamefont {A.}~\bibnamefont {Wang}},\ }\href {\doibase
		10.1103/PhysRevD.110.064044} {\bibfield  {journal} {\bibinfo  {journal}
			{Phys. Rev. D}\ }\textbf {\bibinfo {volume} {110}},\ \bibinfo {pages}
		{064044} (\bibinfo {year} {2024})},\ \Eprint
	{http://arxiv.org/abs/2304.09025} {arXiv:2304.09025 [gr-qc]} \BibitemShut
	{NoStop}%
	\bibitem [{\citenamefont {Liang}\ \emph {et~al.}(2017)\citenamefont {Liang},
		\citenamefont {Gong}, \citenamefont {Hou},\ and\ \citenamefont
		{Liu}}]{Liang:2017ahj}%
	\BibitemOpen
	\bibfield  {author} {\bibinfo {author} {\bibfnamefont {D.}~\bibnamefont
			{Liang}}, \bibinfo {author} {\bibfnamefont {Y.}~\bibnamefont {Gong}},
		\bibinfo {author} {\bibfnamefont {S.}~\bibnamefont {Hou}}, \ and\ \bibinfo
		{author} {\bibfnamefont {Y.}~\bibnamefont {Liu}},\ }\href {\doibase
		10.1103/PhysRevD.95.104034} {\bibfield  {journal} {\bibinfo  {journal} {Phys.
				Rev. D}\ }\textbf {\bibinfo {volume} {95}},\ \bibinfo {pages} {104034}
		(\bibinfo {year} {2017})},\ \Eprint {http://arxiv.org/abs/1701.05998}
	{arXiv:1701.05998 [gr-qc]} \BibitemShut {NoStop}%
	\bibitem [{\citenamefont {Hou}\ \emph {et~al.}(2018)\citenamefont {Hou},
		\citenamefont {Gong},\ and\ \citenamefont {Liu}}]{Hou:2017bqj}%
	\BibitemOpen
	\bibfield  {author} {\bibinfo {author} {\bibfnamefont {S.}~\bibnamefont
			{Hou}}, \bibinfo {author} {\bibfnamefont {Y.}~\bibnamefont {Gong}}, \ and\
		\bibinfo {author} {\bibfnamefont {Y.}~\bibnamefont {Liu}},\ }\href {\doibase
		10.1140/epjc/s10052-018-5869-y} {\bibfield  {journal} {\bibinfo  {journal}
			{Eur. Phys. J. C}\ }\textbf {\bibinfo {volume} {78}},\ \bibinfo {pages} {378}
		(\bibinfo {year} {2018})},\ \Eprint {http://arxiv.org/abs/1704.01899}
	{arXiv:1704.01899 [gr-qc]} \BibitemShut {NoStop}%
	\bibitem [{\citenamefont {Soudi}\ \emph {et~al.}(2019)\citenamefont {Soudi},
		\citenamefont {Farrugia}, \citenamefont {Gakis}, \citenamefont {Levi~Said},\
		and\ \citenamefont {Saridakis}}]{Soudi:2018dhv}%
	\BibitemOpen
	\bibfield  {author} {\bibinfo {author} {\bibfnamefont {I.}~\bibnamefont
			{Soudi}}, \bibinfo {author} {\bibfnamefont {G.}~\bibnamefont {Farrugia}},
		\bibinfo {author} {\bibfnamefont {V.}~\bibnamefont {Gakis}}, \bibinfo
		{author} {\bibfnamefont {J.}~\bibnamefont {Levi~Said}}, \ and\ \bibinfo
		{author} {\bibfnamefont {E.~N.}\ \bibnamefont {Saridakis}},\ }\href {\doibase
		10.1103/PhysRevD.100.044008} {\bibfield  {journal} {\bibinfo  {journal}
			{Phys. Rev. D}\ }\textbf {\bibinfo {volume} {100}},\ \bibinfo {pages}
		{044008} (\bibinfo {year} {2019})},\ \Eprint
	{http://arxiv.org/abs/1810.08220} {arXiv:1810.08220 [gr-qc]} \BibitemShut
	{NoStop}%
	\bibitem [{\citenamefont {Gong}\ \emph {et~al.}(2018)\citenamefont {Gong},
		\citenamefont {Hou}, \citenamefont {Liang},\ and\ \citenamefont
		{Papantonopoulos}}]{Gong:2018cgj}%
	\BibitemOpen
	\bibfield  {author} {\bibinfo {author} {\bibfnamefont {Y.}~\bibnamefont
			{Gong}}, \bibinfo {author} {\bibfnamefont {S.}~\bibnamefont {Hou}}, \bibinfo
		{author} {\bibfnamefont {D.}~\bibnamefont {Liang}}, \ and\ \bibinfo {author}
		{\bibfnamefont {E.}~\bibnamefont {Papantonopoulos}},\ }\href {\doibase
		10.1103/PhysRevD.97.084040} {\bibfield  {journal} {\bibinfo  {journal} {Phys.
				Rev. D}\ }\textbf {\bibinfo {volume} {97}},\ \bibinfo {pages} {084040}
		(\bibinfo {year} {2018})},\ \Eprint {http://arxiv.org/abs/1801.03382}
	{arXiv:1801.03382 [gr-qc]} \BibitemShut {NoStop}%
	\bibitem [{\citenamefont {Capozziello}\ \emph {et~al.}(2020)\citenamefont
		{Capozziello}, \citenamefont {Capriolo},\ and\ \citenamefont
		{Caso}}]{Capozziello:2019msc}%
	\BibitemOpen
	\bibfield  {author} {\bibinfo {author} {\bibfnamefont {S.}~\bibnamefont
			{Capozziello}}, \bibinfo {author} {\bibfnamefont {M.}~\bibnamefont
			{Capriolo}}, \ and\ \bibinfo {author} {\bibfnamefont {L.}~\bibnamefont
			{Caso}},\ }\href {\doibase 10.1140/epjc/s10052-020-7737-9} {\bibfield
		{journal} {\bibinfo  {journal} {Eur. Phys. J. C}\ }\textbf {\bibinfo {volume}
			{80}},\ \bibinfo {pages} {156} (\bibinfo {year} {2020})},\ \Eprint
	{http://arxiv.org/abs/1912.12469} {arXiv:1912.12469 [gr-qc]} \BibitemShut
	{NoStop}%
	\bibitem [{\citenamefont {Liang}\ \emph {et~al.}(2022)\citenamefont {Liang},
		\citenamefont {Xu}, \citenamefont {Lu},\ and\ \citenamefont
		{Shao}}]{Liang:2022hxd}%
	\BibitemOpen
	\bibfield  {author} {\bibinfo {author} {\bibfnamefont {D.}~\bibnamefont
			{Liang}}, \bibinfo {author} {\bibfnamefont {R.}~\bibnamefont {Xu}}, \bibinfo
		{author} {\bibfnamefont {X.}~\bibnamefont {Lu}}, \ and\ \bibinfo {author}
		{\bibfnamefont {L.}~\bibnamefont {Shao}},\ }\href {\doibase
		10.1103/PhysRevD.106.124019} {\bibfield  {journal} {\bibinfo  {journal}
			{Phys. Rev. D}\ }\textbf {\bibinfo {volume} {106}},\ \bibinfo {pages}
		{124019} (\bibinfo {year} {2022})},\ \Eprint
	{http://arxiv.org/abs/2207.14423} {arXiv:2207.14423 [gr-qc]} \BibitemShut
	{NoStop}%
	\bibitem [{\citenamefont {Dong}\ \emph {et~al.}(2024)\citenamefont {Dong},
		\citenamefont {Liu},\ and\ \citenamefont {Liu}}]{Dong:2023bgt}%
	\BibitemOpen
	\bibfield  {author} {\bibinfo {author} {\bibfnamefont {Y.-Q.}\ \bibnamefont
			{Dong}}, \bibinfo {author} {\bibfnamefont {Y.-Q.}\ \bibnamefont {Liu}}, \
		and\ \bibinfo {author} {\bibfnamefont {Y.-X.}\ \bibnamefont {Liu}},\ }\href
	{\doibase 10.1103/PhysRevD.109.044013} {\bibfield  {journal} {\bibinfo
			{journal} {Phys. Rev. D}\ }\textbf {\bibinfo {volume} {109}},\ \bibinfo
		{pages} {044013} (\bibinfo {year} {2024})},\ \Eprint
	{http://arxiv.org/abs/2310.11336} {arXiv:2310.11336 [gr-qc]} \BibitemShut
	{NoStop}%
	\bibitem [{\citenamefont {Dong}\ \emph {et~al.}(2025)\citenamefont {Dong},
		\citenamefont {Lai}, \citenamefont {Liu},\ and\ \citenamefont
		{Liu}}]{Dong:2024zal}%
	\BibitemOpen
	\bibfield  {author} {\bibinfo {author} {\bibfnamefont {Y.-Q.}\ \bibnamefont
			{Dong}}, \bibinfo {author} {\bibfnamefont {X.-B.}\ \bibnamefont {Lai}},
		\bibinfo {author} {\bibfnamefont {Y.-Q.}\ \bibnamefont {Liu}}, \ and\
		\bibinfo {author} {\bibfnamefont {Y.-X.}\ \bibnamefont {Liu}},\ }\href
	{\doibase 10.1140/epjc/s10052-025-14378-5} {\bibfield  {journal} {\bibinfo
			{journal} {Eur. Phys. J. C}\ }\textbf {\bibinfo {volume} {85}},\ \bibinfo
		{pages} {645} (\bibinfo {year} {2025})},\ \Eprint
	{http://arxiv.org/abs/2409.11838} {arXiv:2409.11838 [gr-qc]} \BibitemShut
	{NoStop}%
	\bibitem [{\citenamefont {Cornish}\ and\ \citenamefont
		{Littenberg}(2015)}]{Cornish:2014kda}%
	\BibitemOpen
	\bibfield  {author} {\bibinfo {author} {\bibfnamefont {N.~J.}\ \bibnamefont
			{Cornish}}\ and\ \bibinfo {author} {\bibfnamefont {T.~B.}\ \bibnamefont
			{Littenberg}},\ }\href {\doibase 10.1088/0264-9381/32/13/135012} {\bibfield
		{journal} {\bibinfo  {journal} {Class. Quant. Grav.}\ }\textbf {\bibinfo
			{volume} {32}},\ \bibinfo {pages} {135012} (\bibinfo {year} {2015})},\
	\Eprint {http://arxiv.org/abs/1410.3835} {arXiv:1410.3835 [gr-qc]}
	\BibitemShut {NoStop}%
	\bibitem [{\citenamefont {Abbott}\ \emph {et~al.}(2016)\citenamefont {Abbott}
		\emph {et~al.}}]{LIGOScientific:2016lio}%
	\BibitemOpen
	\bibfield  {author} {\bibinfo {author} {\bibfnamefont {B.~P.}\ \bibnamefont
			{Abbott}} \emph {et~al.} (\bibinfo {collaboration} {LIGO Scientific,
			Virgo}),\ }\href {\doibase 10.1103/PhysRevLett.116.221101} {\bibfield
		{journal} {\bibinfo  {journal} {Phys. Rev. Lett.}\ }\textbf {\bibinfo
			{volume} {116}},\ \bibinfo {pages} {221101} (\bibinfo {year} {2016})},\
	\bibinfo {note} {[Erratum: Phys.Rev.Lett. 121, 129902 (2018)]},\ \Eprint
	{http://arxiv.org/abs/1602.03841} {arXiv:1602.03841 [gr-qc]} \BibitemShut
	{NoStop}%
	\bibitem [{\citenamefont {Abbott}\ \emph
		{et~al.}(2019{\natexlab{b}})\citenamefont {Abbott} \emph
		{et~al.}}]{LIGOScientific:2019fpa}%
	\BibitemOpen
	\bibfield  {author} {\bibinfo {author} {\bibfnamefont {B.~P.}\ \bibnamefont
			{Abbott}} \emph {et~al.} (\bibinfo {collaboration} {LIGO Scientific,
			Virgo}),\ }\href {\doibase 10.1103/PhysRevD.100.104036} {\bibfield  {journal}
		{\bibinfo  {journal} {Phys. Rev. D}\ }\textbf {\bibinfo {volume} {100}},\
		\bibinfo {pages} {104036} (\bibinfo {year} {2019}{\natexlab{b}})},\ \Eprint
	{http://arxiv.org/abs/1903.04467} {arXiv:1903.04467 [gr-qc]} \BibitemShut
	{NoStop}%
	\bibitem [{\citenamefont {Abbott}\ \emph
		{et~al.}(2020{\natexlab{b}})\citenamefont {Abbott} \emph
		{et~al.}}]{LIGOScientific:2020zkf}%
	\BibitemOpen
	\bibfield  {author} {\bibinfo {author} {\bibfnamefont {R.}~\bibnamefont
			{Abbott}} \emph {et~al.} (\bibinfo {collaboration} {LIGO Scientific,
			Virgo}),\ }\href {\doibase 10.3847/2041-8213/ab960f} {\bibfield  {journal}
		{\bibinfo  {journal} {Astrophys. J. Lett.}\ }\textbf {\bibinfo {volume}
			{896}},\ \bibinfo {pages} {L44} (\bibinfo {year} {2020}{\natexlab{b}})},\
	\Eprint {http://arxiv.org/abs/2006.12611} {arXiv:2006.12611 [astro-ph.HE]}
	\BibitemShut {NoStop}%
	\bibitem [{\citenamefont {Abbott}\ \emph
		{et~al.}(2020{\natexlab{c}})\citenamefont {Abbott} \emph
		{et~al.}}]{LIGOScientific:2020ufj}%
	\BibitemOpen
	\bibfield  {author} {\bibinfo {author} {\bibfnamefont {R.}~\bibnamefont
			{Abbott}} \emph {et~al.} (\bibinfo {collaboration} {LIGO Scientific,
			Virgo}),\ }\href {\doibase 10.3847/2041-8213/aba493} {\bibfield  {journal}
		{\bibinfo  {journal} {Astrophys. J. Lett.}\ }\textbf {\bibinfo {volume}
			{900}},\ \bibinfo {pages} {L13} (\bibinfo {year} {2020}{\natexlab{c}})},\
	\Eprint {http://arxiv.org/abs/2009.01190} {arXiv:2009.01190 [astro-ph.HE]}
	\BibitemShut {NoStop}%
	\bibitem [{\citenamefont {Abbott}\ \emph
		{et~al.}(2021{\natexlab{b}})\citenamefont {Abbott} \emph
		{et~al.}}]{LIGOScientific:2020tif}%
	\BibitemOpen
	\bibfield  {author} {\bibinfo {author} {\bibfnamefont {R.}~\bibnamefont
			{Abbott}} \emph {et~al.} (\bibinfo {collaboration} {LIGO Scientific,
			Virgo}),\ }\href {\doibase 10.1103/PhysRevD.103.122002} {\bibfield  {journal}
		{\bibinfo  {journal} {Phys. Rev. D}\ }\textbf {\bibinfo {volume} {103}},\
		\bibinfo {pages} {122002} (\bibinfo {year} {2021}{\natexlab{b}})},\ \Eprint
	{http://arxiv.org/abs/2010.14529} {arXiv:2010.14529 [gr-qc]} \BibitemShut
	{NoStop}%
	\bibitem [{\citenamefont {Abbott}\ \emph {et~al.}(2025)\citenamefont {Abbott}
		\emph {et~al.}}]{LIGOScientific:2021sio}%
	\BibitemOpen
	\bibfield  {author} {\bibinfo {author} {\bibfnamefont {R.}~\bibnamefont
			{Abbott}} \emph {et~al.} (\bibinfo {collaboration} {LIGO Scientific, VIRGO,
			KAGRA}),\ }\href {\doibase 10.1103/PhysRevD.112.084080} {\bibfield  {journal}
		{\bibinfo  {journal} {Phys. Rev. D}\ }\textbf {\bibinfo {volume} {112}},\
		\bibinfo {pages} {084080} (\bibinfo {year} {2025})},\ \Eprint
	{http://arxiv.org/abs/2112.06861} {arXiv:2112.06861 [gr-qc]} \BibitemShut
	{NoStop}%
	\bibitem [{\citenamefont {Green}\ and\ \citenamefont
		{Moffat}(2018)}]{Green:2017voq}%
	\BibitemOpen
	\bibfield  {author} {\bibinfo {author} {\bibfnamefont {M.~A.}\ \bibnamefont
			{Green}}\ and\ \bibinfo {author} {\bibfnamefont {J.~W.}\ \bibnamefont
			{Moffat}},\ }\href {\doibase 10.1016/j.physletb.2018.08.009} {\bibfield
		{journal} {\bibinfo  {journal} {Phys. Lett. B}\ }\textbf {\bibinfo {volume}
			{784}},\ \bibinfo {pages} {312} (\bibinfo {year} {2018})},\ \Eprint
	{http://arxiv.org/abs/1711.00347} {arXiv:1711.00347 [astro-ph.IM]}
	\BibitemShut {NoStop}%
	\bibitem [{\citenamefont {Nielsen}\ \emph {et~al.}(2019)\citenamefont
		{Nielsen}, \citenamefont {Nitz}, \citenamefont {Capano},\ and\ \citenamefont
		{Brown}}]{Nielsen:2018bhc}%
	\BibitemOpen
	\bibfield  {author} {\bibinfo {author} {\bibfnamefont {A.~B.}\ \bibnamefont
			{Nielsen}}, \bibinfo {author} {\bibfnamefont {A.~H.}\ \bibnamefont {Nitz}},
		\bibinfo {author} {\bibfnamefont {C.~D.}\ \bibnamefont {Capano}}, \ and\
		\bibinfo {author} {\bibfnamefont {D.~A.}\ \bibnamefont {Brown}},\ }\href
	{\doibase 10.1088/1475-7516/2019/02/019} {\bibfield  {journal} {\bibinfo
			{journal} {JCAP}\ }\textbf {\bibinfo {volume} {02}},\ \bibinfo {pages} {019}
		(\bibinfo {year} {2019})},\ \Eprint {http://arxiv.org/abs/1811.04071}
	{arXiv:1811.04071 [astro-ph.HE]} \BibitemShut {NoStop}%
	\bibitem [{\citenamefont {Marcoccia}\ \emph {et~al.}(2020)\citenamefont
		{Marcoccia}, \citenamefont {Fredriksson}, \citenamefont {Nielsen},\ and\
		\citenamefont {Nardini}}]{Marcoccia:2020rag}%
	\BibitemOpen
	\bibfield  {author} {\bibinfo {author} {\bibfnamefont {P.}~\bibnamefont
			{Marcoccia}}, \bibinfo {author} {\bibfnamefont {F.}~\bibnamefont
			{Fredriksson}}, \bibinfo {author} {\bibfnamefont {A.~B.}\ \bibnamefont
			{Nielsen}}, \ and\ \bibinfo {author} {\bibfnamefont {G.}~\bibnamefont
			{Nardini}},\ }\href {\doibase 10.1088/1475-7516/2020/11/043} {\bibfield
		{journal} {\bibinfo  {journal} {JCAP}\ }\textbf {\bibinfo {volume} {11}},\
		\bibinfo {pages} {043} (\bibinfo {year} {2020})},\ \Eprint
	{http://arxiv.org/abs/2008.12663} {arXiv:2008.12663 [gr-qc]} \BibitemShut
	{NoStop}%
	\bibitem [{\citenamefont {Liang}\ \emph {et~al.}(2020)\citenamefont {Liang},
		\citenamefont {Kanner}, \citenamefont {Weinstein}, \citenamefont {Ghonge},\
		and\ \citenamefont {Gong}}]{Liang:2020rt}%
	\BibitemOpen
	\bibfield  {author} {\bibinfo {author} {\bibfnamefont {D.}~\bibnamefont
			{Liang}}, \bibinfo {author} {\bibfnamefont {J.}~\bibnamefont {Kanner}},
		\bibinfo {author} {\bibfnamefont {A.~J.}\ \bibnamefont {Weinstein}}, \bibinfo
		{author} {\bibfnamefont {S.}~\bibnamefont {Ghonge}}, \ and\ \bibinfo {author}
		{\bibfnamefont {Y.}~\bibnamefont {Gong}},\ }\href@noop {} {\enquote {\bibinfo
			{title} {{Residuals for Binary Black Hole Events in GWTC-1 Using Chi-squared
					Tests}},}\ } (\bibinfo {year} {2020}),\ \bibinfo {note}
	{{LIGO-P2000464}}\BibitemShut {NoStop}%
	\bibitem [{\citenamefont {Chatterji}\ \emph {et~al.}(2004)\citenamefont
		{Chatterji}, \citenamefont {Blackburn}, \citenamefont {Martin},\ and\
		\citenamefont {Katsavounidis}}]{Chatterji:2004qg}%
	\BibitemOpen
	\bibfield  {author} {\bibinfo {author} {\bibfnamefont {S.}~\bibnamefont
			{Chatterji}}, \bibinfo {author} {\bibfnamefont {L.}~\bibnamefont
			{Blackburn}}, \bibinfo {author} {\bibfnamefont {G.}~\bibnamefont {Martin}}, \
		and\ \bibinfo {author} {\bibfnamefont {E.}~\bibnamefont {Katsavounidis}},\
	}\href {\doibase 10.1088/0264-9381/21/20/024} {\bibfield  {journal} {\bibinfo
			{journal} {Class. Quant. Grav.}\ }\textbf {\bibinfo {volume} {21}},\
		\bibinfo {pages} {S1809} (\bibinfo {year} {2004})},\ \Eprint
	{http://arxiv.org/abs/gr-qc/0412119} {arXiv:gr-qc/0412119} \BibitemShut
	{NoStop}%
	\bibitem [{\citenamefont {Chatterji}(2005)}]{chatterji2005search}%
	\BibitemOpen
	\bibfield  {author} {\bibinfo {author} {\bibfnamefont {S.~K.}\ \bibnamefont
			{Chatterji}},\ }\emph {\bibinfo {title} {The search for gravitational wave
			bursts in data from the second LIGO science run}},\ \href@noop {} {Ph.D.
		thesis},\ \bibinfo  {school} {Massachusetts Institute of Technology}
	(\bibinfo {year} {2005})\BibitemShut {NoStop}%
	\bibitem [{\citenamefont {Blackburn}\ \emph {et~al.}(2008)\citenamefont
		{Blackburn} \emph {et~al.}}]{Blackburn:2008ah}%
	\BibitemOpen
	\bibfield  {author} {\bibinfo {author} {\bibfnamefont {L.}~\bibnamefont
			{Blackburn}} \emph {et~al.},\ }\href {\doibase
		10.1088/0264-9381/25/18/184004} {\bibfield  {journal} {\bibinfo  {journal}
			{Class. Quant. Grav.}\ }\textbf {\bibinfo {volume} {25}},\ \bibinfo {pages}
		{184004} (\bibinfo {year} {2008})},\ \Eprint {http://arxiv.org/abs/0804.0800}
	{arXiv:0804.0800 [gr-qc]} \BibitemShut {NoStop}%
	\bibitem [{\citenamefont {Vazsonyi}\ and\ \citenamefont
		{Davis}(2023)}]{Vazsonyi:2022jul}%
	\BibitemOpen
	\bibfield  {author} {\bibinfo {author} {\bibfnamefont {L.}~\bibnamefont
			{Vazsonyi}}\ and\ \bibinfo {author} {\bibfnamefont {D.}~\bibnamefont
			{Davis}},\ }\href {\doibase 10.1088/1361-6382/acafd2} {\bibfield  {journal}
		{\bibinfo  {journal} {Class. Quant. Grav.}\ }\textbf {\bibinfo {volume}
			{40}},\ \bibinfo {pages} {035008} (\bibinfo {year} {2023})},\ \Eprint
	{http://arxiv.org/abs/2208.12338} {arXiv:2208.12338 [astro-ph.IM]}
	\BibitemShut {NoStop}%
	\bibitem [{\citenamefont {Aasi}\ \emph {et~al.}(2015)\citenamefont {Aasi} \emph
		{et~al.}}]{LIGOScientific:2014pky}%
	\BibitemOpen
	\bibfield  {author} {\bibinfo {author} {\bibfnamefont {J.}~\bibnamefont
			{Aasi}} \emph {et~al.} (\bibinfo {collaboration} {LIGO Scientific}),\ }\href
	{\doibase 10.1088/0264-9381/32/7/074001} {\bibfield  {journal} {\bibinfo
			{journal} {Class. Quant. Grav.}\ }\textbf {\bibinfo {volume} {32}},\ \bibinfo
		{pages} {074001} (\bibinfo {year} {2015})},\ \Eprint
	{http://arxiv.org/abs/1411.4547} {arXiv:1411.4547 [gr-qc]} \BibitemShut
	{NoStop}%
	\bibitem [{\citenamefont {Trovato}(2020)}]{Trovato:2019liz}%
	\BibitemOpen
	\bibfield  {author} {\bibinfo {author} {\bibfnamefont {A.}~\bibnamefont
			{Trovato}} (\bibinfo {collaboration} {Ligo Scientific, Virgo}),\ }\href
	{\doibase 10.22323/1.357.0082} {\bibfield  {journal} {\bibinfo  {journal}
			{PoS}\ }\textbf {\bibinfo {volume} {Asterics2019}},\ \bibinfo {pages} {082}
		(\bibinfo {year} {2020})}\BibitemShut {NoStop}%
	\bibitem [{\citenamefont {Acernese}\ \emph {et~al.}(2015)\citenamefont
		{Acernese} \emph {et~al.}}]{VIRGO:2014yos}%
	\BibitemOpen
	\bibfield  {author} {\bibinfo {author} {\bibfnamefont {F.}~\bibnamefont
			{Acernese}} \emph {et~al.} (\bibinfo {collaboration} {VIRGO}),\ }\href
	{\doibase 10.1088/0264-9381/32/2/024001} {\bibfield  {journal} {\bibinfo
			{journal} {Class. Quant. Grav.}\ }\textbf {\bibinfo {volume} {32}},\ \bibinfo
		{pages} {024001} (\bibinfo {year} {2015})},\ \Eprint
	{http://arxiv.org/abs/1408.3978} {arXiv:1408.3978 [gr-qc]} \BibitemShut
	{NoStop}%
	\bibitem [{\citenamefont {Pratten}\ \emph {et~al.}(2021)\citenamefont {Pratten}
		\emph {et~al.}}]{Pratten:2020ceb}%
	\BibitemOpen
	\bibfield  {author} {\bibinfo {author} {\bibfnamefont {G.}~\bibnamefont
			{Pratten}} \emph {et~al.},\ }\href {\doibase 10.1103/PhysRevD.103.104056}
	{\bibfield  {journal} {\bibinfo  {journal} {Phys. Rev. D}\ }\textbf {\bibinfo
			{volume} {103}},\ \bibinfo {pages} {104056} (\bibinfo {year} {2021})},\
	\Eprint {http://arxiv.org/abs/2004.06503} {arXiv:2004.06503 [gr-qc]}
	\BibitemShut {NoStop}%
	\bibitem [{\citenamefont {Macleod}\ \emph {et~al.}(2021)\citenamefont
		{Macleod}, \citenamefont {Areeda}, \citenamefont {Coughlin}, \citenamefont
		{Massinger},\ and\ \citenamefont {Urban}}]{Macleod:2021goi}%
	\BibitemOpen
	\bibfield  {author} {\bibinfo {author} {\bibfnamefont {D.~M.}\ \bibnamefont
			{Macleod}}, \bibinfo {author} {\bibfnamefont {J.~S.}\ \bibnamefont {Areeda}},
		\bibinfo {author} {\bibfnamefont {S.~B.}\ \bibnamefont {Coughlin}}, \bibinfo
		{author} {\bibfnamefont {T.~J.}\ \bibnamefont {Massinger}}, \ and\ \bibinfo
		{author} {\bibfnamefont {A.~L.}\ \bibnamefont {Urban}},\ }\href {\doibase
		10.1016/j.softx.2021.100657} {\bibfield  {journal} {\bibinfo  {journal}
			{SoftwareX}\ }\textbf {\bibinfo {volume} {13}},\ \bibinfo {pages} {100657}
		(\bibinfo {year} {2021})}\BibitemShut {NoStop}%
	\bibitem [{\citenamefont {An}(1933)}]{an1933sulla}%
	\BibitemOpen
	\bibfield  {author} {\bibinfo {author} {\bibfnamefont {K.}~\bibnamefont
			{An}},\ }\href@noop {} {\bibfield  {journal} {\bibinfo  {journal} {Giorn
				Dell'inst Ital Degli Att}\ }\textbf {\bibinfo {volume} {4}},\ \bibinfo
		{pages} {89} (\bibinfo {year} {1933})}\BibitemShut {NoStop}%
	\bibitem [{\citenamefont {Smirnov}(1948)}]{smirnov1948table}%
	\BibitemOpen
	\bibfield  {author} {\bibinfo {author} {\bibfnamefont {N.}~\bibnamefont
			{Smirnov}},\ }\href@noop {} {\bibfield  {journal} {\bibinfo  {journal} {The
				annals of mathematical statistics}\ }\textbf {\bibinfo {volume} {19}},\
		\bibinfo {pages} {279} (\bibinfo {year} {1948})}\BibitemShut {NoStop}%
	\bibitem [{\citenamefont {Virtanen}\ \emph {et~al.}(2020)\citenamefont
		{Virtanen}, \citenamefont {Gommers}, \citenamefont {Oliphant}, \citenamefont
		{Haberland}, \citenamefont {Reddy}, \citenamefont {Cournapeau}, \citenamefont
		{Burovski}, \citenamefont {Peterson}, \citenamefont {Weckesser},
		\citenamefont {Bright} \emph {et~al.}}]{virtanen2020scipy}%
	\BibitemOpen
	\bibfield  {author} {\bibinfo {author} {\bibfnamefont {P.}~\bibnamefont
			{Virtanen}}, \bibinfo {author} {\bibfnamefont {R.}~\bibnamefont {Gommers}},
		\bibinfo {author} {\bibfnamefont {T.~E.}\ \bibnamefont {Oliphant}}, \bibinfo
		{author} {\bibfnamefont {M.}~\bibnamefont {Haberland}}, \bibinfo {author}
		{\bibfnamefont {T.}~\bibnamefont {Reddy}}, \bibinfo {author} {\bibfnamefont
			{D.}~\bibnamefont {Cournapeau}}, \bibinfo {author} {\bibfnamefont
			{E.}~\bibnamefont {Burovski}}, \bibinfo {author} {\bibfnamefont
			{P.}~\bibnamefont {Peterson}}, \bibinfo {author} {\bibfnamefont
			{W.}~\bibnamefont {Weckesser}}, \bibinfo {author} {\bibfnamefont
			{J.}~\bibnamefont {Bright}},  \emph {et~al.},\ }\href@noop {} {\bibfield
		{journal} {\bibinfo  {journal} {Nature methods}\ }\textbf {\bibinfo {volume}
			{17}},\ \bibinfo {pages} {261} (\bibinfo {year} {2020})}\BibitemShut
	{NoStop}%
	\bibitem [{\citenamefont {Anderson}\ and\ \citenamefont
		{Darling}(1954)}]{anderson1954test}%
	\BibitemOpen
	\bibfield  {author} {\bibinfo {author} {\bibfnamefont {T.~W.}\ \bibnamefont
			{Anderson}}\ and\ \bibinfo {author} {\bibfnamefont {D.~A.}\ \bibnamefont
			{Darling}},\ }\href@noop {} {\bibfield  {journal} {\bibinfo  {journal}
			{Journal of the American statistical association}\ }\textbf {\bibinfo
			{volume} {49}},\ \bibinfo {pages} {765} (\bibinfo {year} {1954})}\BibitemShut
	{NoStop}%
	\bibitem [{\citenamefont {Pearson}(1900)}]{pearson1900x}%
	\BibitemOpen
	\bibfield  {author} {\bibinfo {author} {\bibfnamefont {K.}~\bibnamefont
			{Pearson}},\ }\href@noop {} {\bibfield  {journal} {\bibinfo  {journal} {The
				London, Edinburgh, and Dublin Philosophical Magazine and Journal of Science}\
		}\textbf {\bibinfo {volume} {50}},\ \bibinfo {pages} {157} (\bibinfo {year}
		{1900})}\BibitemShut {NoStop}%
	\bibitem [{\citenamefont {Punturo}\ \emph {et~al.}(2010)\citenamefont {Punturo}
		\emph {et~al.}}]{Punturo:2010zz}%
	\BibitemOpen
	\bibfield  {author} {\bibinfo {author} {\bibfnamefont {M.}~\bibnamefont
			{Punturo}} \emph {et~al.},\ }\href {\doibase 10.1088/0264-9381/27/19/194002}
	{\bibfield  {journal} {\bibinfo  {journal} {Class. Quant. Grav.}\ }\textbf
		{\bibinfo {volume} {27}},\ \bibinfo {pages} {194002} (\bibinfo {year}
		{2010})}\BibitemShut {NoStop}%
	\bibitem [{\citenamefont {Reitze}\ \emph {et~al.}(2019)\citenamefont {Reitze}
		\emph {et~al.}}]{Reitze:2019iox}%
	\BibitemOpen
	\bibfield  {author} {\bibinfo {author} {\bibfnamefont {D.}~\bibnamefont
			{Reitze}} \emph {et~al.},\ }\href@noop {} {\bibfield  {journal} {\bibinfo
			{journal} {Bull. Am. Astron. Soc.}\ }\textbf {\bibinfo {volume} {51}},\
		\bibinfo {pages} {035} (\bibinfo {year} {2019})},\ \Eprint
	{http://arxiv.org/abs/1907.04833} {arXiv:1907.04833 [astro-ph.IM]}
	\BibitemShut {NoStop}%
	\bibitem [{\citenamefont {Hild}\ \emph {et~al.}(2011)\citenamefont {Hild} \emph
		{et~al.}}]{Hild:2010id}%
	\BibitemOpen
	\bibfield  {author} {\bibinfo {author} {\bibfnamefont {S.}~\bibnamefont
			{Hild}} \emph {et~al.},\ }\href {\doibase 10.1088/0264-9381/28/9/094013}
	{\bibfield  {journal} {\bibinfo  {journal} {Class. Quant. Grav.}\ }\textbf
		{\bibinfo {volume} {28}},\ \bibinfo {pages} {094013} (\bibinfo {year}
		{2011})},\ \Eprint {http://arxiv.org/abs/1012.0908} {arXiv:1012.0908 [gr-qc]}
	\BibitemShut {NoStop}%
\end{thebibliography}

%

\end{document}